\documentclass[sigconf,screen]{acmart}
\usepackage{graphicx}
\usepackage{amssymb}
\usepackage{bbding}
\usepackage{tikz}
\usepackage{tabularx}
\usepackage{multirow}
\usepackage{multicol}
\usetikzlibrary{trees}
\usepackage{longtable}
\usepackage{makecell}
\usepackage{booktabs}
\usepackage{threeparttable} % 用于更专业的表格注释
\usepackage{subfig}
\AtBeginDocument{%
  \providecommand\BibTeX{{%
    \normalfont B\kern-0.5em{\scshape i\kern-0.25em b}\kern-0.8em\TeX}}}

\copyrightyear{2026}
\acmYear{2026}
\setcopyright{cc}
\setcctype{by}
\acmConference[CHI '26]{Proceedings of the 2026 CHI Conference on Human Factors in Computing Systems}{April 13--17, 2026}{Barcelona, Spain}
\acmBooktitle{Proceedings of the 2026 CHI Conference on Human Factors in Computing Systems (CHI '26), April 13--17, 2026, Barcelona, Spain}
\acmPrice{}
\acmDOI{10.1145/3772318.3790455}
\acmISBN{979-8-4007-2278-3/2026/04}

\begin{document}

\title[Examining Perceived Privacy Risk Across Data Transmission and Sharing Ranges of SPA]{``Privacy across the boundary'': Examining Perceived Privacy Risk Across Data Transmission and Sharing Ranges of Smart Home Personal Assistants}

\author{Shuning Zhang}
\orcid{0000-0002-4145-117X}
\affiliation{%
  \institution{Tsinghua University}
  \city{Beijing}
  \country{China}
}
\email{zsn23@mails.tsinghua.edu.cn}

\author{Shixuan Li}
\orcid{0009-0008-6828-6347}
\affiliation{%
  \institution{Tsinghua University}
  \city{Beijing}
  \country{China}
}
\email{li-sx24@mails.tsinghua.edu.cn}

\author{Haobin Xing}
\orcid{0009-0007-0492-4613}
\affiliation{%
  \institution{Tsinghua University}
  \city{Beijing}
  \country{China}
}
\email{xinghb21@mails.tsinghua.edu.cn}

\author{Jiarui Liu}
\orcid{0009-0007-8937-7128}
\affiliation{%
  \institution{Tsinghua University}
  \city{Beijing}
  \country{China}
}
\email{15774169531@163.com}

\author{Yan Kong}
\orcid{0000-0002-4187-3559}
\affiliation{%
  \institution{Tsinghua University}
  \city{Beijing}
  \country{China}
}
\email{ky21@mails.tsinghua.edu.cn}

\author{Xin Yi}
\orcid{0000-0001-8041-7962}
\authornote{Corresponding author.}
\affiliation{
    \institution{Tsinghua University}
    \city{Beijing}
    \country{China}
}
\affiliation{
    \institution{Beijing Academy of Artificial Intelligence}
    \city{Beijing}
    \country{China}
}
\email{yixin@tsinghua.edu.cn}

\author{Kanye Ye Wang}
\orcid{0000-0001-7908-5286}
\affiliation{
    \institution{University of Macau}
    \city{Macao}
    \country{China}
}
\email{wangye@um.edu.mo}

\author{Hewu Li}
\orcid{0000-0002-6331-6542}
\affiliation{
    \institution{Tsinghua University}
    \city{Beijing}
    \country{China}
}
\email{lihewu@cernet.edu.cn}

\renewcommand{\shortauthors}{Zhang et al.}


\begin{abstract}
As Smart Home Personal Assistants (SPAs) evolve into social agents, understanding user privacy necessitates interpersonal communication frameworks, such as Privacy Boundary Theory (PBT). To ground our investigation, our three-phase preliminary study (1) identified transmission and sharing ranges as key boundary-related risk factors, (2) categorized relevant SPA functions and data types, and (3) analyzed commercial practices, revealing widespread data sharing and non-transparent safeguards. A subsequent mixed-methods study (N=412 survey, N=40 interviews among the survey participants) assessed users' perceived privacy risks across data types, transmission ranges and sharing ranges. Results demonstrate a significant, non-linear escalation in perceived risk when data crosses two critical boundaries: the `public network' (transmission) and `third parties' (sharing). This boundary effect holds robustly across data types and demographics. Furthermore, risk perception is modulated by data attributes (e.g., social relational data), and contextual privacy calculus. Conversely, anonymization safeguards show limited efficacy especially for third-party sharing, a finding attributed to user distrust. These findings empirically ground PBT in the SPA context and inform design of boundary-aware privacy protection.
\end{abstract}




\maketitle

\section{Introduction}

The proliferation of Smart Home Personal Assistants (SPAs) (e.g., Xiaodu\footnote{https://smarthome.baidu.com/advantage}, Alexa\footnote{https://www.alexa.com/})~\cite{edu2020smart,abdi2021privacy} marks a significant evolution in human-computer interaction, with a global user base exceeding 422 million\footnote{\url{https://www.coolest-gadgets.com/smart-home-devices-statistics/}} and 68 million in China alone by 2024\footnote{\url{https://www.sohu.com/a/795255808_121752158}}~\cite{edu2020smart,sciuto2018hey}. As these devices integrate deeply into daily life and connect with vast ecosystems of third-party services (e.g., Spotify, Uber), they caused substantial concerns of privacy that necessitate targeted investigation~\cite{edu2020smart}. Prior research has primarily focused on technical vulnerabilities~\cite{zhang2025sok,fernandes2016security,haack2017security}, mitigation strategies~\cite{mcmillan2019designing,mhaidli2020listen,chen2020wearable}, or specific data channels~\cite{dubois2020speakers,denning2013computer}, often adopting a system security of functional data-flow perspective~\cite{fruchter2018consumer,lau2018alexa,abdi2021privacy}. 

However, as SPAs increasingly integrate anthropomorphic features~\cite{schneiders2021effect}, the user-SPA relationship transcends simple tool interaction~\cite{edu2020smart}, instead resembling engagement with a social agent~\cite{luria2016designing,oh2024better}. This shift from an inanimate tool to an interactive agent renders traditional privacy models insufficient, particularly those that frame privacy management as a binary (on/off) or static (disclose/not disclose) choice~\cite{felt2012android,felt2012ask}. Consequently, privacy management with an SPA evolves from configuring a single setting~\cite{felt2012android,felt2012ask} to a nuanced, continuous negotiation analogous to that found in interpersonal communication~\cite{petronio1991communication,petronio2002boundaries}. When a device operates as an agent, the user's management strategy becomes more complex, necessitating re-modeling of their mental model rather than simple, one-time permissions.

This dynamic mandates theoretical frameworks grounded in interpersonal communication to interpret user privacy management. Privacy Boundary Theory (PBT)~\cite{petronio2002boundaries,petronio2010communication}, originally focusing on interpersonal communication, provides such a framework. PBT explains how individuals, conceptualized as ``information owners'', manage access to private information by establishing and \textit{\textbf{regulating privacy boundaries}}, potentially engaging entities like SPAs as ``co-owners'' of disclosed information. These boundaries are managed using rules influenced by \textit{\textbf{attributes of the information, context, demographics and privacy calculus}}~\cite{petronio1991communication,petronio2002boundaries}. As SPAs increasingly function as social agents~\cite{oh2024better}, it becomes crucial to understand how users navigate interactions by managing boundaries based on perceived personal and social consequences~\cite{strauss2013social} associated with different data types.

Therefore, this paper leverages PBT to investigate users' perceived privacy risks within the SPA ecosystem, focusing on two fundamental aspects. First, acknowledging PBT's emphasis on boundary structures, we examine the impact of \textit{boundary-related factors} on users' perceived privacy risks. These factors are defined as those that constitute privacy boundaries, and influence users' perceived privacy risks~\cite{guo2025not}. Second, recognizing PBT's principle of boundary coordination, we investigate how privacy boundaries vary across \textit{boundary regulation-related factors}, such as \textit{demographics}, and \textit{privacy calculus}. These factors govern how users manage and adjust their boundaries across contexts. Notably, we focus on perceived privacy risk as it reveals users' mental models when user manage trust and adoption~\cite{zeng2017end,zheng2018user}. While stated perceptions can diverge from actual behavior (i.e., the privacy paradox~\cite{kokolakis2017privacy}), understanding this perceptual framework is a critical and under-explored precursor to interpreting user actions and designing systems that align with their expectations. These considerations lead to our research questions (RQs):
% that boundary management rules depend on attributes of the information, demographics, contexts and privacy calculus, we examine how privacy boundaries further vary with these \textbf{boundary regulation-related factors} Note that we differentiate boundary-related factors from boundary regulation-related factors in that the former ones defined boundaries while the latter ones modulated boundaries. Notably, we focus on perceptions as they form the foundational mental models users employ to manage trust and assess violations~\cite{zeng2017end,zheng2018user}. While stated perceptions can diverge from actual behavior (i.e., the privacy paradox~\cite{kokolakis2017privacy}), understanding this perceptual framework is a critical and under-explored precursor to interpreting user actions and designing systems that align with their expectations. These considerations lead to our research questions (RQs):

\noindent \textbf{RQ1: [Privacy Boundary Effect]} What are the effects of privacy boundaries (instantiated as boundary-related factors) on users' perceived privacy risk of SPAs?

\noindent \textbf{RQ2: [Variance of Privacy Boundary]} How do privacy boundaries and perceived privacy risks vary across different boundary regulation-related factors?

Towards these RQs, our preliminary study conducted focus groups to identify ``transmission'' and ``sharing'' ranges as key boundary-related factors, and performed a synthesis to categorize 23 relevant SPA functions. Subsequently, we conducted a large-scale survey (N=412) and follow-up interviews (N=40), applying these boundary-related factors to assess privacy boundaries and perceived privacy risks across various data categories. The choice of Chinese users was motivated by their large and growing SPA user base, which however has been previously under-explored~\cite{edu2020smart,abdi2021privacy,abdi2019more}.

For RQ1, our findings reveal that users' perceived privacy risk is significantly escalated by crossing privacy boundaries of transmission range (managing spatial boundaries) and sharing range (managing relational boundaries). Specifically, for transmission, we identified the boundary when data traverse the home perimeter compared to within the home resulted in a significant escalation in perceived risk. This boundary effect was consistent across diverse data categories, underscoring the home boundary's role as a primary cognitive heuristic tied to spatial sanctuary. Concerning sharing, we identified the boundary when data crosses the provider to third-party threshold resulted in a marked increase in risk, although privacy risk is still perceived even for within-provider sharing, signifying a perceived loss of control and breach of relational trust when data moves beyond the provider.

For RQ2, we systematically investigated how key boundary regulation-related factors modulate perceived privacy risk. \textbf{First}, risk perception varied significantly with attributes of the private data. Users highlighted heightened sensitivity towards inherently personal data (e.g., communications) versus operational data, reflecting subjective `data sensitivity boundaries'. \textbf{Second}, user awareness of data practices and the perceived functional necessity of the data collection also significantly correlated with perceived risk levels, factoring into their overall privacy risk calculus. \textbf{Third}, users' demographics had a  limited effect on privacy boundaries, confirming the broad applicability of fundamental boundary rules across different user groups. \textbf{Last}, technical safeguards, especially anonymization on third-party sharing, showed limited effectiveness in mitigating boundary effects. This was largely attributed to user skepticism regarding actual implementation and inherent distrust when unknown entities are involved. Collectively, these findings underscore the multi-faceted and context-dependent nature of privacy boundary regulation. To summarize, the contributions of this paper are three-fold:

$\bullet$ [Theoretical] We extend PBT to SPAs by establishing the ``Home Network'' and ``Service Provider'' as critical privacy boundaries, where risk escalates non-linearly due to violations of information co-ownership.

$\bullet$ [Empirical] We examined the impact of information types and technical safeguards on the effect of boundaries, revealing that mechanisms like anonymization fail to mitigate risk because users' privacy calculus is dominated by relational trust.

$\bullet$ [Implications] We propose boundary-aware design implications, with hierarchical visualizations and threshold-triggered alerts to align protections with users' boundaries.

\section{Backgrounds \& Related Work}

We survey PBT for insights into how users manage information boundaries when interacting with SPAs, and review prior work detailing user concerns about SPA, thereby contextualizing our analysis of SPA privacy perceptions.

\begin{table*}[t]
    \centering
    \caption{Comparison of our work and prior literature (Quan: Quantitative, Qual: Qualitative, CI: Contextual Integrity Theory, CL: Cognitive Load Theory, PBT: Privacy Boundary Theory).}
    \label{tab:comparison}
    \small
    % 使用 tabularx 环境，宽度设为 \textwidth
    % @{} 移除两侧多余的空格
    % l：左对齐；c：居中；X：自动宽度并左对齐长文本内容
    \begin{tabularx}{\textwidth}{@{} p{0.5cm} c c p{4cm} c p{6cm} @{}} 
        \toprule
        % 使用 \textbf 加粗表头
        \textbf{Ref} & \textbf{Year} & \textbf{\makecell{Qual/Quan}} & \textbf{Factors} & \textbf{Theory} & \textbf{Core Finding} \\
        \midrule
        
        \cite{fruchter2018consumer} & 2018 & Quan & Reviews & --- & SPA has creepy behavior \\
        \cite{lau2018alexa} & 2018 & Qual & --- & --- & Users trade privacy for utility \\
        \cite{apthorpe2018discovering} & 2018 & Quan & Sender, Recipient, Attribute, Transmission principle & CI & Contextual norms \\
        \cite{abdi2019more} & 2019 & Qual & --- & CI & Users have incomplete threat models \\
        \cite{cho2019hey} & 2019 & Quan & Modality/device, Information sensitivity & CL & Modality impact users' attitude of SPAs \\
        \cite{malkin2019privacy} & 2019 & Mixed & --- & CI & Users have incomplete threat models \\
        \cite{cho2020will} & 2020 & Quan & Customization or not & --- & Customize privacy settings enhance trust and usability \\
        \cite{edu2020smart} & 2020 & --- & --- & CI & --- \\
        \cite{huang2020amazon} & 2020 & Qual & --- & --- & Users have incomplete understandings \\
        \cite{abdi2021privacy} & 2021 & Quan & Recipient, Purpose, Data type, Transmission Principle & CI & Privacy norms and contexual rules \\
        \cite{sabir2022hey} & 2022 & Qual & --- & --- & Users can hardly identify skill developer correctly \\
        
        % \midrule % 使用 \midrule 明确分隔 prior work 和 our work
        
        \textbf{Ours} & \textbf{--} & \textbf{Quan} & \textbf{Transmission ranges, Sharing ranges} & \textbf{PBT} & \textbf{Non-linear boundaries when crossing the home and service providers} \\
        \bottomrule
    \end{tabularx}
\end{table*}

\subsection{Privacy Boundary Theory}\label{sec:rw_contextual}

\textbf{Theoretical Foundations.} PBT~\cite{petronio2002boundaries}, originally named as Communication Privacy Management (CPM) theory~\cite{petronio1991communication,altman1975environment}, provides a framework for understanding how individuals manage the tension between revealing and concealing private information. In its evolution, Petronio~\cite{petronio2004road} traces the shift of PBT from early self-disclosure research to a rule-based management system, positing that individuals perceived private information as their property. To manage this ``ownership'', individuals set privacy boundaries encompassing the self and authorized others. When information is shared, the recipient becomes a ``co-owner'', necessitating the coordination of privacy rules regarding permeability (how much is shared), linkage (who else has access), and control (rights to further dissemination)~\cite{petronio2010managing,sannon2020just}. 

These privacy rules~\cite{dickerson2016communication,henningsen2019student,hernandez2018understanding,smith2017reveal} are guided by several criteria: (1) cultural norms, which establish general expectations for information sharing, (2) individual characteristics and motivations, such as relationship development versus self-protection, (3) attributes of the private information, including its perceived sensitivity, (4) contextual factors, such as the specific setting (e.g., home versus work) and the mediating role of technical measures, and (5) risk-benefit assessments, where individuals weigh the perceived advantages of disclosure against potential negative consequences~\cite{liu2018regulate,petronio2002boundaries}. Notably, this management process is dynamic. When expectations are violated, ``boundary turbulence'' occurs, prompting individuals to recalibrate their privacy rules to restore control~\cite{petronio2004road,child2015privacy}. 

\textbf{PBT in Human-Computer Interaction (HCI) and Human-AI Interaction (HAI).} Given the increasingly social nature of the SPA ecosystem, where human interacts with smart assistants, we examine the evolution of PBT in HCI and HAI. While originally used for interpersonal communication~\cite{petronio2002boundaries}, PBT has been extensively adapted to Computer-Mediated Communication (CMC) and social media within the HCI community, where boundaries are increasingly permeable and collective. Research indicates that digital privacy is often managed in layers (individual vs. group) and complicates boundary coordination due to the collapse of physical context~\cite{mansour2021collective,child2010blogging}. Cho et al.~\cite{cho2018collective} further demonstrates that the causal pathways from PBT theories to self-disclosure are culturally variable, validating the theory's applicability across diverse populations (e.g., US, Singapore, South Korea).

Recent HCI work has introduced nuanced refinements to PBT, such as ``privacy silence''--a non-verbal boundary setting facilitated by mobile devices in intimate relationships~\cite{amirkhani2025privacy}--and the specific challenges faced by neurodiverse populations, where restrictive mediation by support networks can induce boundary turbulence regarding digital independence~\cite{cullen2024towards}. Furthermore, research in algorithmic environments identifies that factors like ``privacy fatigue'' and algorithm awareness significantly influence the adoption of open privacy management practices, challenging the view of passive user resignation~\cite{choi2018role}. 

As interactions shift from human-mediated to AI-mediated contexts, HCI scholars have extended PBT to HAI, treating the artificial agent as a non-human co-owner of information. Sannon et al.~\cite{sannon2020just} found that unauthorized data sharing by conversational chatbots is also viewed as a privacy violation similar to human sharing, particularly when data is linked to advertisers. Similarly, Liu et al.~\cite{liu2024chatbots} found that agent identity and information sensitivity serve as control rules that dictate boundary linkage and permeability, with users manifesting different behavioral intentions towards chatbots versus humans. In immersive environments like Social Virtual Reality (SVR), Cummings~\cite{cummings2025self} notes that perceived actorhood and non-permeable boundaries are essential for fostering self-disclosure. Beyond these perceptual factors, the practical delegation of boundary regulation to AI systems remains complex. Guo et al.~\cite{guo2025not} propose AI-powered elicitation methods to capture personal boundaries, yet note that user willingness to delegate varies significantly by context. \textbf{While prior work treats agents primarily as a social interface, we differed by also considering the boundary complexities inherent in SPA's ecosystem. Furthermore, moving beyond qualitative examinations of agent identity, we adopt a quantitative focus on measuring and validating these specific boundary effects (e.g., spatial and relational).}

\textbf{Operationalizing Privacy Boundaries in SPAs.} In the specific domain of smart speakers, prior work has primarily focused on general attitudinal pathways, such as precautionary versus protective motivations~\cite{xu2022smart}. We differentiate our work by two aspects. First, distinct from the focus on general trust, we examine how data flows and the ecosystem functions as determinants of privacy boundaries. Second, moving beyond general attitudes, we measure concrete boundary thresholds operationalized by perceived privacy risks. This approach identifies specific systemic boundaries to inform boundary-aware design.
% While prior studies assess trust and attitudes, they often overlook how the ecosystem and data flow serves as the influential factors for privacy boundaries. Besides, beyond the attitudinal focus, we aim to measure the concrete boundaries defined by perceived privacy risks to inform designs.

In human-SPA interactions, we conceptualize the user as the information owner and the SPA as an authorized co-owner, establishing privacy boundaries around information exchanges~\cite{petronio2002boundaries}. Following established practices that correlate boundary regulation with data flows~\cite{cho2018collective,sannon2020just} and quantify boundaries via risk perception~\cite{guo2025not}, we employ perceived privacy risk as the primary metric for boundary instantiation. Given the importance of \textit{boundary permeability} and \textit{boundary linkage} in privacy boundaries' construct~\cite{petronio2002boundaries}, we map the regulation of \textit{boundary permeability} and \textit{boundary linkage} separately to \textit{transmission ranges} (defining spatial boundary) and \textit{sharing ranges} (defining relational boundary), a framework derived from our preliminary study. 

\subsection{Data Collection and Privacy Concerns of SPAs}

The evolution of personal assistants from functional command interfaces to anthropomorphic social agents~\cite{oh2024better,schneiders2021effect}, especially in home contexts, represents a paradigm shift in Human-Computer Interaction (HCI). As these devices emulate social presence and integrate deeply into intimate spaces~\cite{shi2025sociable,zhang2022she}, user-device relationship transcends simple tool utilization, resembling engagement with a social counterpart~\cite{ha2021exploring}. The shift fosters examination from the perspective of boundary regulations, exemplified by works focusing on everyday intelligent agents~\cite{sannon2020just,singh2023between,xie2025you}. Specific to SPAs, to examine how users navigate this landscape, we synthesize the evolution of prior work from qualitative to quantitative into four streams: (1) \textbf{mental models}, where technical opacity hinders understanding, (2) \textbf{privacy calculus}, where social and functional benefits trade-off against risk, (3) \textbf{interaction dynamics}, where multi-user contexts and anthropomorphism obscure risk perception, and (4) \textbf{ecosystem norms}, where third-party complexities challenge contextual integrity.

\textbf{Regarding user mental models and awareness}, a significant body of work highlights that users possess incomplete or erroneous understandings of SPA data practices. Early research by Zeng et al.~\cite{zeng2017end} and Lau et al.~\cite{lau2018alexa} identified that while users express abstract privacy concerns, they often lack specific technical understanding, leading to ad hoc mitigation strategies and a reliance on misplaced trust. This lack of transparency extends to data storage and processing. Abdi et al.~\cite{abdi2019more} found that users' understanding is often limited to the physical household, failing to comprehend cloud processing or third-party involvement. Furthermore, specific data practices remain opaque. For instance, Malkin et al.~\cite{malkin2019privacy} revealed that many users were unaware that voice recordings were permanently stored or reviewable. Consequently, users often underestimate their responsibility in the shared security model~\cite{haney2021s}, adopting suboptimal management strategies due to limited options or technical knowledge~\cite{haney2020smart,huang2020amazon}. Unlike these studies which primarily identify the gaps in user understanding, our work adopts a granular perspective to investigate how users perceived privacy risks with different data transmission and sharing ranges, utilizing PBT to explain the nature of these perceptions.

\textbf{Regarding privacy calculus~\cite{laufer1977privacy},} past literature suggests that SPA adoption is driven by a trade-off where perceived benefits often override privacy concerns. Studies indicate that users generally tolerate privacy risks in exchange for convenience, connectedness, and utility~\cite{manikonda2018s,zheng2018user,vimalkumar2021okay}. Wang et al.~\cite{wang2018want} utilized a net valence model to show that users tend to ignore potential risks to prioritize performance expectancy. However, this acceptance is fragile. Cho et al.~\cite{cho2020will} found that while customization enhances trust, it can backfire among privacy-conscious users if not paired with content control. Additionally, Meng et al.~\cite{meng2021owning} noted that despite high adoption, concerns persist regarding constant monitoring and lack of transparency. While prior work frames risk as a general variable weighted against benefit, our work advances this by identifying specific boundary crossings as critical factors that non-linearly escalated the perceived risks.

\textbf{Regarding multi-user and modality context,} these dynamics introduce structural complexities to risk perception. Research has shown that bystanders often possess different privacy concerns than owners but lack control mechanisms~\cite{alshehri2022exploring,zeng2019understanding}. Furthermore, the anthropomorphic nature of voice interaction can enhance social presence and positive attitudes, potentially masking privacy risks~\cite{cho2019hey}. While acknowledging these dynamics, our current study isolates the structural impact of transmission and sharing boundaries on the primary user's risk perception, providing a basis for analyzing the escalation of privacy risks.

\textbf{Regarding norms and data flows,} research grounded in Contextual Integrity (CI) framework has mapped appropriate information flows in smart homes. CI posits that privacy is preserved when information flows adhere to context-specific norms defined by five distinct parameters: sender, recipient, subject, information type, and transmission principle~\cite{nissenbaum2004privacy}. Applying this lens, Apthorpe et al.~\cite{apthorpe2018discovering} and Abdi et al.~\cite{abdi2021privacy} have mapped privacy norms, identifying that users' acceptance of data collection is highly dependent on contextual parameters such as the data recipient and purpose~\cite{lee2016understanding,barbosa2019if}. Tabassum et al.~\cite{tabassum2019investigating} further refined this by showing that consent is tied to data sensitivity and perceived service benefit. However, the expanding third-party ecosystem complicated these norms. Edu et al.~\cite{edu2020smart} noted that research has focused narrowly on user-device interaction, often overlooking the broad attack surface. Crucially, users struggle to distinguish between native and third-party skills~\cite{major2021alexa,sabir2022hey}, often failing to identify skill developers correctly. \textbf{We distinguish our work from CI-based approaches~\cite{apthorpe2018discovering,abdi2021privacy} by contributing a quantitative validation of boundary-based risk structures across data types.}

\section{Preliminary Study}

To ground the privacy boundary investigation, we first identified \textbf{boundary-related} factors on perceived privacy risks via focus groups. Crucially, this preliminary phase aimed solely to isolate the factors that \textit{define} privacy boundaries, rather than eliciting regulation factors that \textit{modulate} boundaries. We then systematically categorized SPA functions and sensitive data types, and finally contextualized these by analyzing current commercial practices. These results informed the subsequent study's scope and methodology.

\subsection{Phase 1: Investigating Boundary-related Factors Influencing Perceived Privacy Risks towards SPAs via Focus Groups}\label{sec:investigation}

\paragraph{Methodology}
To investigate boundary-related factors shaping user perceptions of SPA privacy risks, we conducted focus groups~\cite{moju2022you,morgan1996focus,powell1996focus}. We recruited 21 participants (mean age=22.8, SD=2.3; 13 male, 8 female) via Chinese online platforms, continuing recruitment until reaching theoretical saturation~\cite{morse1995significance}. We first acquired all participants' contact information, and then grouped, contacted, and conducted focus group sessions with participants iteratively. Specifically, the sessions ceased when two consecutive focus group yielded no new themes (i.e., boundary-related factors), totaling 6 focus groups (3--5 participants per group). All participants owned at least one SPA, qualifying them as `early adopters'~\cite{rogers2014diffusion,zheng2018user}. While this targeted sampling strategy facilitated focused exploration, we acknowledge that recruitment exclusively via Chinese online platforms and the sample's composition of young adult `early adopters' potentially restrict the generalizability of findings across diverse cultural contexts, age demographics, and user experience levels. 

Before the session, we obtained informed consent and provided standardized definitions for SPAs, privacy risk, and contextual data~\cite{nissenbaum2004privacy} (details in Appendix~\ref{sec:interview_definition}). Each session began with participants individually brainstorming influential factors (20 min). Subsequently, participants within each group shared, deduplicated, and collectively discussed the identified factors for practicality. Sessions lasted approximately 50 minutes on average. We recorded and transcribed the data for analysis. Each participant received 100 RMB as compensation, consistent with local wage standards. The study's protocol was approved by our institution's Institutional Review Board (IRB).

We performed a qualitative analysis using Nvivo on the transcribed data, integrating deductive approaches guided by PBT~\cite{petronio2002boundaries}, with inductive coding~\cite{corbin2014basics}. Two researchers independently coded an initial 10\% sample (2 people) using both deductive and inductive coding techniques~\cite{charmaz2014constructing}, resolving any discrepancies through discussions. For the remaining 90\% of the data, we employed constant comparative analysis~\cite{boeije2002purposeful}. Through axial coding~\cite{strauss1987qualitative}, we differentiated between themes aligned with existing theories and newly emergent themes. In line with established interpretive research standards, we determined that calculating Inter-Rater Reliability (IRR) was inappropriate for this qualitative analysis~\cite{mcdonald2019reliability}.

\begin{figure}[!htbp]
    \centering
    \includegraphics[width=0.5\textwidth]{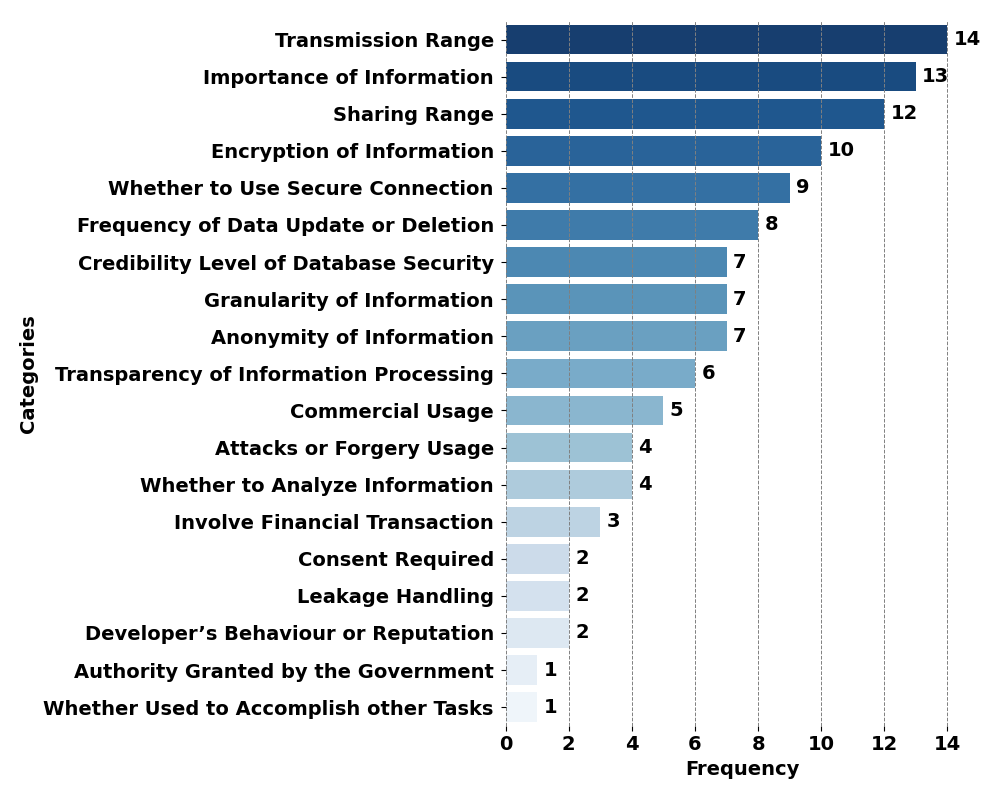}
    \caption{The factors users mentioned (arranged from the most mentioned to the least mentioned). The number indicated the frequency.}
    \label{fig:spa_study1}
\end{figure}

\paragraph{Results}\label{sec:brainstorm_result}

Our analysis identified three key factors as predominantly influencing user perceptions of SPA privacy risks (see Figure~\ref{fig:spa_study1}): \textbf{transmission range} (mentioned 14 times), \textbf{sensitivity of information} (13 times), and \textbf{sharing range} (12 times). They were mentioned by over half of the participants. Specifically, concerns over \textbf{transmission range} map onto the management of spatial or technical boundaries of information flow, \textbf{sharing range} mentions the negotiation of relational boundaries dictating who appropriately receives the information~\cite{petronio2002boundaries}, and \textbf{information sensitivity} highlights the boundaries delineated by the information itself.

Supporting this finding, participants frequently articulated concerns about the scope of data transmission, specifically whether data remained on the local device, was confined to the smart home network, or traversed the broad internet. Similarly, they emphasized the importance of sharing boundaries, expressing particular concern about third-party access. As P2 stated, a key consideration was \textit{``whether the information was disclosed to the broad internet or shared with third-party companies.''} Sensitivity concerns predominantly focused on physiological data or other forms of personally identifiable information. The classifications participants used for sharing range (local device, service provider, third-party entities) mirrored the hierarchy observed for transmission range.

We noted that participants often conflated technical specifics, expressing unease about external data flows despite sometimes lacking a precise understanding of network topologies or data handling processes, similar to prior work's findings~\cite{zeng2017end}. This led several participants to advocate for localized processing, with P6 suggesting, \textit{``These personal information could be processed just inside the device.''} However, others acknowledged existing technical limitations, as P10 observed, \textit{``edge computing has not yet enabled that complicated functions.''} While participants mentioned additional factors like data granularity and anonymity, they did so less frequently. These findings collectively underscore users' focus on boundary control and transparency in data handling, highlighting the inherent tension between the functional benefits of SPAs and the associated privacy risks within current technological ecosystems.

\subsection{Phase 2: Determining SPA Functions and Data Collection}\label{sec:determine_function}

\paragraph{Methodology}

This phase employed a two-step methodology, collection with initial categorization, followed by filtering, to identify relevant SPA functions and associated data practices. 

1. Collection and Initial Categorization: We first compiled a comprehensive inventory of potential SPA functions and the data they collect. This involved synthesizing taxonomies from prior research~\cite{edu2020smart,abdi2019more,lau2018alexa,alhadlaq1902privacy,abdi2021privacy} and examining the features of prominent commercial SPAs, namely Amazon Alexa\footnote{\url{https://www.amazon.com/alexa-skills/}} and Google Assistant\footnote{\url{https://support.google.com/assistant/answer/7172842}}. Data sources included academic literature~\cite{abdi2019more,abdi2021privacy,lau2018alexa,edu2020smart,zeng2019understanding} and publicly available product documentation\footnote{\url{https://en.wikipedia.org/wiki/Amazon_Alexa}} on online websites. Concurrently, to structure the analysis, we performed an initial functional grouping. Inspired by common SPA usage patterns and literature~\cite{edu2020smart,abdi2021privacy,lau2018alexa}, we preliminarily grouped functions based on their primary intended purpose or functional domain (e.g., health, contact, search). Two primary researchers performed this collection and categorization, coding each function and associated data. They identified discrepancies in coding and resolved them through focused discussion until consensus was reached. 

2. Filtering: We selected the functions that necessitate involvement from service providers or third-party companies, as these represented key areas where boundary management is crucial. Functions primarily involving local device control or built-in operations were excluded. The finalized functional categorization was confirmed by the entire research team after the filtering stage.

\paragraph{Results}\label{sec:spa_function_result}

1. Collection and Initial Categorization: The initial collection phase yielded 64 unique SPA functions and 183 function-data item pairs, encompassing 34 distinct data types. These functions were classified into seven groups: personal basic functions (7), multimedia entertainment functions (7), device control functions
(22), health-related functions (6), contact-related functions (5), consumption-related functions (2), and search-related functions (15). Among these, the most frequently collected data types were `preference' (required by 35 functions), `location' (15), and `account data' (15). The complete initial list is detailed in Table~\ref{usage_spa}. 

2. Filtering: we selected 23 functions and 72 associated function-data item pairs, representing 21 unique data types for further analysis (detailed in Table~\ref{tbl:information_collection}). Based on the functional grouping performed during collection and refined after filtering, these 23 selected functions were mapped into five core functional categories relevant to external data sharing: \textit{multimedia entertainment}, \textit{health}, \textit{contact}, \textit{consumption}, and \textit{search}. Functions initially grouped under `personal basic function' (7 functions) and `device control' (22 functions) were excluded during filtering as they did not meet our criterion of requiring external service provider involvement. The detailed mapping of the final selected functions and data items to these categories is presented in Tables~\ref{usage_spa} and~\ref{tbl:information_collection}.

\subsection{Phase 3: Real-world Implementation Investigation}

\paragraph{Methodology}

To ground participants' concerns, we collected and analyzed the data transmission and sharing practices of prominent commercial products according to the list from wikipedia\footnote{\url{https://en.wikipedia.org/wiki/Smart_speaker}, experiment conducted on Sep 2024}. Our analysis was performed with publicly searchable websites, including official documents and unofficial websites. We acknowledge this approach relies on publicly available documentation and may not capture undisclosed or dynamically changing practices. The products we selected included Amazon Alexa, Bixby, Google Assistant, etc., totaling 14 types (Details could be seen in Table~\ref{fig:merged_table} in Appendix~\ref{app:real_world}). We aggregated the SPAs to the service provider levels, as providers manage service delivery and data collection. We searched official websites for important aspects identified in Section~\ref{sec:investigation}, including data transmission, sharing, encryption, and anonymization practices, following previous guidelines~\cite{bitaab2023beyond}. This process mirrored how users become familiar with their devices (while acknowledging stated policies may not fully capture implementation). We classified the implementation of these practices into four categories: Unclear, Part, No and All. ``Unclear'' denotes the privacy policy did not indicate whether to collect the corresponding information. ``Part'' denotes the privacy policy states that part of the information would be collected and shared. ``No'' denotes that no information would be collected, shared or transmitted in this manner. ``All'' denotes that all information would be collected, shared or transmitted in this manner. 

\paragraph{Results}\label{sec:industry_results}

Our analysis found that data transmission and sharing are widespread and reveal a fundamental lack of user privacy control, which provided the ground to contrast with users' perceptions in Sections~\ref{sec:rq1} and~\ref{sec:rq2} (See Tables~\ref{fig:merged_table} and~\ref{real_world_info} in Appendix~\ref{app:real_world} for more details). 

\textbf{1. Data transmission and third-party sharing are widespread.} In our real world investigation, 6/10 of SPAs reported transmitting user data to the broader internet, and 5/10 reported sharing all collected data with third-party services. In contrast, only 1/10 explicitly state that they do not share data with third-party services. This widespread transmission and sharing underscore a pervasive lack of privacy control, which may pose significant risks to users. 

\textbf{2. Encryption and anonymization are often unclear or inadequately communicated.} Half of the SPAs (5/10) did not specify whether they encrypt user data, and another 5/10 failed to clarify their anonymization practices. This lack of transparency regarding encryption and anonymization may expose users to potential data breaches, suggesting the need for clearer and more consistent security measures across the industry. 

\textbf{3. The range of data collected by SPAs is extensive, with significant variation across products.} Nearly all SPAs (9/10) reported collecting a broad array of user information, including location data, device usage, and voice recordings. However, for more sensitive data such as user comments, posts, and gift card information, the collection is more limited. Only 2/10 of SPAs reported collecting user comments or posts, and similarly, only 2/10 reported collecting gift card details. This highlights the inconsistency in data collection practices, with some service providers collecting a wide range of information while others limit their data collection to less sensitive categories. Service providers exhibit diverse data collection practices. 

\textbf{4. The extent and scope of data collection vary greatly across service providers.} For instance, while nearly all products reported collecting user names, email addresses, and device usage data, some products (e.g., Mycroft and Ding Dong) do not report collecting user social media information or biometric data. This diversity suggests that reported data collection practices are not standardized, creating an opportunity for regulation and alignment across providers.

\section{Main Study: Examining The Effect of Transmission and Sharing Ranges on Users' Perceived Privacy Risks of SPAs}

To answer the RQs, we conducted a study to investigate how users' privacy risk perceptions of SPAs are influenced by data type, transmission range and sharing. Through large-scale surveys and interviews, we examined the effects of encryption, anonymization, user awareness and perceived data importance. 

\subsection{Study Design and Procedure}\label{sec:study_design_procedure}

This study investigated how users perceive privacy risks associated with SPAs based on three primary factors: \textit{data type}, \textit{transmission range} and \textit{sharing range}. These factors were selected based on our preliminary studies (Sections~\ref{sec:brainstorm_result} and ~\ref{sec:spa_function_result}, see also Figure~\ref{fig:spa_study1}). Besides, we also examined the effect of \textit{anonymization} and \textit{encryption} on transmission and sharing ranges separately, as the perceived risks of transmission and sharing may be influenced by the users' mental models regarding anonymity and encryption~\cite{tabassum2019don}. 

The transmission range involved three levels: \textit{on-device processing,} where data is exclusively on the SPA device itself, without being transmitted to any external location; \textit{in-home transmission,} where data is transmitted from the SPA to other devices or a local server only within the users' home network; \textit{internet (public) transmission,} where data is transmitted from the SPA to external servers (e.g., the service provider's cloud) across the broad internet. The sharing range also involved three levels: \textit{No external sharing (on device),} where data is confined to the SPA device and not shared with any external entity; \textit{within the device's provider,} where data is shared with the primary service provider of the SPA, or \textit{with third-party companies,} where data is shared with companies other than the primary service provider. 

We further examined scenarios incorporating safeguards: data \textit{transmitted to the internet with encryption}, data \textit{shared with the device's provider with anonymization}, and data \textit{shared with third-party companies with anonymization}. For all these conditions, participants assessed the associated privacy risks. We also collected data on participants' \textit{awareness of the data practice} and the \textit{perceived importance of the data for device functioning} to explore their privacy calculus.

The survey began with an introduction to SPAs, their functions, the types of data collected, and the meaning of each entry. We also used vignettes to facilitate users' understanding (Appendix~\ref{sec:content_questionnaire}). Participants were presented with function categories (e.g., multimedia entertainment, health) and rated 10 entries per function. We conducted three rounds of pilot studies to iterate and refine our questionnaire. All entries were measured using 7-point Likert scales, with 1 being least risky and 7 most risky, following prior practice~\cite{bhatia2018empirical} (details of questionnaire justifications and pilot studies see Appendix~\ref{sec:justification}).

To manage the extensive number of data types (72 in total), we randomly selected 35-40 data types for each participant, ensuring even distribution of responses through a Python algorithm using the shuffle function. Participants were given five 90-second breaks instantiated as blank times blocks to reduce fatigue, and ensure qualified participation~\cite{backor2007estimating}. Following the questionnaire, we randomly selected about 10\% of participants for in-depth semi-structured interviews to explore their privacy risk perceptions~\cite{grober2023cloud}. Example questions included (details see Appendix~\ref{sec:interview_script}):

1. What were your privacy risk ratings for different transmission and sharing ranges, and what factors influence those ratings? [cause]

2. What concerns did you have regarding these transmission and sharing ranges, and why did certain factors increase your perceived privacy risk? [factor]

3. How do you think anonymization and encryption could mitigate privacy risks? [mediation]

4. In your opinion, how do different functions affect perceived privacy risks, and why? [function]

\begin{figure}[!htbp]
    \includegraphics[width=0.47\textwidth]{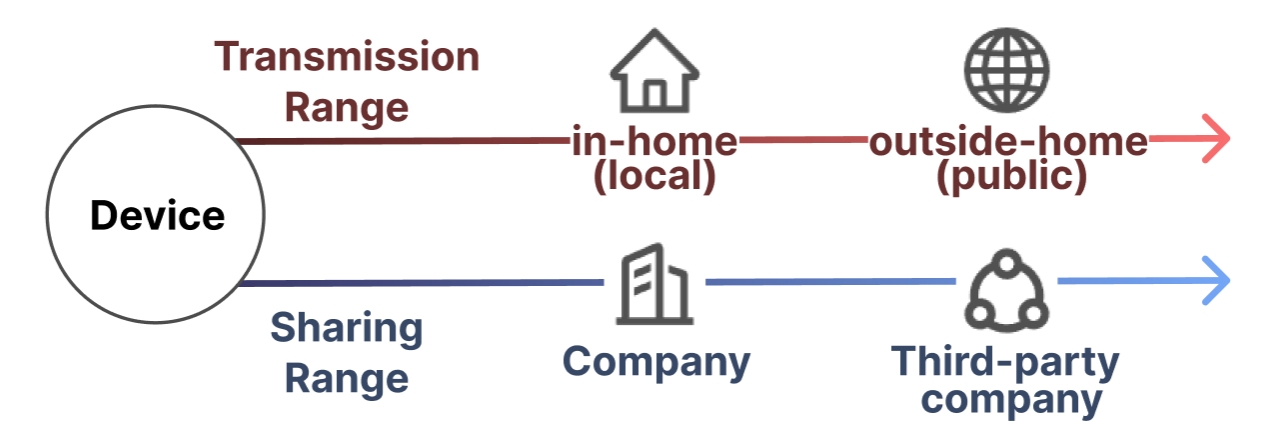}
    \caption{Different factor levels of transmission and sharing range, where company is usually the service provider.}
    \label{fig:transmis_share}
\end{figure}

\subsection{Participants}

The IRB-approved study was conducted in China using an online questionnaire\footnote{distribution platform: \url{http://www.wjx.cn}}. The survey and interview were conducted using Chinese. Participants were required to be over 18 years old, as enforced by the preset on the website, with no additional restrictions to ensure the inclusion of diverse demographics (e.g., both current and potential SPA users). We finally recruited 412 participants (190 males, 221 females, 1 prefer not to say) with diverse educational backgrounds, ages, and professions. Most participants were aged 26-35 (243/412) or 36-45 (93/412) and held a bachelor’s degree (328/412). The majority of participants were with mechanics/fabrication, finance/commerce or IT related degree or occupations, counting 146, 87 and 79 each (recruitment details see Appendix~\ref{sec:study2_recruitment}). Participants received 35 RMB for completing the questionnaire according to the local wage standard. We selected 40 participants in the interview according to criteria outlined in Section~\ref{sec:study_design_procedure}. This subset participating in follow-up interviews received an additional 100 RMB.

\subsection{Data Analysis}

% We analyzed the quantitative data using statistical methods. Since the data violated the assumption of homogeneity of variances, we employed non-parametric tests, including the Friedman test \cite{pereira2015overview} (for analyzing transmission and sharing ranges) and Kruskal-Wallis test (for analyzing the effects of demographics). Post-hoc comparisons were conducted using the Nemenyi non-parametric test, with Bonferroni correction~\cite{siegel1957nonparametric} applied. Statistical significance was reported for $p < .05$. For the correlation analysis, we performed linear regression with no normalization, as the data were all 7-point Likert scales.
We analyzed the quantitative data using a combination of statistical methods, including Friedman tests~\cite{pereira2015overview}, Kruskal-Wallis tests, corresponding post-hoc tests with Bonferroni corrections~\cite{siegel1957nonparametric} and correlation analyses. We adopted thematic analysis~\cite{braun2019reflecting} on the interview transcripts, where two researchers jointly and iteratively coded the corpus to form and refine codebooks. They solved conflicts and ensured coding quality with intermittent discussions. After coding, researchers aggregated them into themes together. Due to the inductive nature of this coding process, we did not report the results of reliability metrics~\cite{mcdonald2019reliability}.

\subsection{Limitations}

We acknowledge several limitations in the main study. First, regarding cultural and regional context, the study was conducted in China. Consequently, the findings regarding SPA usage and data collection perceptions may not be directly applicable to other regions, particularly given differences in regulatory landscapes (e.g., GDPR~\cite{voigt2017eu}) and the culturally specific nature of privacy~\cite{xu2023dipa}. Future research would benefit from cross-cultural investigations to validate these boundaries globally~\cite{pentina2016exploring}. 

Second, regarding recruitment, participants were sourced via an online platform. These participants may inherently skew towards populations that are more tech-savvy and educated than the general public~\cite{abdi2019more,abdi2021privacy}, however we aimed for a natural distribution across demographics and such platforms are widely validated in privacy research~\cite{abdi2019more,abdi2021privacy}.

Third, our reliance on self-reported data may diverge from actual behaviors due to the privacy paradox~\cite{kokolakis2017privacy}, and recall bias~\cite{lazar2017research}. Our findings highlighted the perceived boundaries, which set the foundation for understanding behavioral deviations, and future research can leverage this baseline to quantify when and why users compromise these preferences in real-world scenarios.

Finally, regarding multi-user and temporal dynamics, our study focuses on individual, static, and perceived boundaries. It does not encompass multi-user dynamics, such as scenarios involving shared family data~\cite{zeng2019understanding}. Furthermore, privacy management is a dynamic process where boundaries shift with experience and habituation~\cite{chalhoub2021did}. Our work serves as a foundational step in uncovering mental models, advocating for future work to address these interpersonal, temporal and behavioral dimensions.

\section{RQ1: Perceived Privacy Risk Unanimously Surged Across Home and Service Provider Boundary}\label{sec:rq1}

Perceived SPA privacy risk escalates significantly across two boundary types grounded in PBT~\cite{petronio2002boundaries}: transmission range (reflecting spatial contexts) and sharing range (reflecting relational contexts). Specifically, we identified that home perimeter is a critical transmission boundary, and the provider to third-party threshold acts as a critical sharing boundary.

\begin{figure*}[!htbp]
    \centering 
    \includegraphics[width=\textwidth]{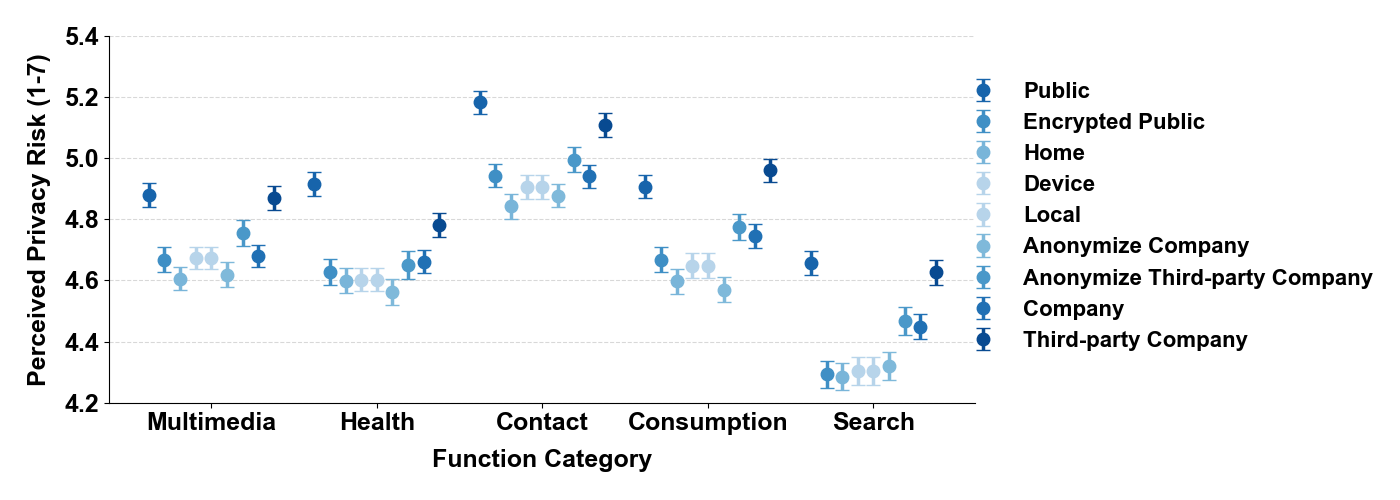}
    \caption{Perceived privacy risks for different transmission ranges and sharing ranges across data categories. 1: lowest risk, 7: highest risk. Errorbar indicated 95\% Confidence Intervals (CIs).}
    \label{fig:transmission_sharing}
\end{figure*}

\subsection{Home Perimeter is a Critical Transmission Boundary} \label{result:transmission}

\textbf{We found that home acts as a critical privacy boundary, where data transmission beyond home triggers a significant increase in perceived privacy risk.} Users perceived similar, relatively low levels of privacy risk for data transmitted locally on the device (M=4.59) and data transmitted within the home (M=4.55). However, perceived risk significantly increased when data was transmitted over the broad internet (M=4.89) (See Figure~\ref{fig:transmission_sharing}). Adopting Friedman's tests with post-hoc Nemenyi tests as the ratings violate normal distribution~\cite{pereira2015overview}, we found this difference was statistically significant, as in Table~\ref{tab:merged_stats}. Post-hoc tests also confirmed the significant increase for broad internet transmission compared to within home conditions. 

\begin{table*}[h]
    \centering
    \caption{Consistency of privacy risk perceptions across boundaries. Friedman tests ($\chi^2$) and Kendall's $W$ indicate consistent, significant effects across all categories and in the aggregate analysis.Interpretation relies on conventional thresholds: $W \ge .7$ indicates very strong agreement, $.5 \le W < .7$ strong, $.3 \le W < .5$ moderate, and $W < .3$ weak.}
    \label{tab:merged_stats}
    % \renewcommand{\arraystretch}{1.1} % 稍微增加行高，提升可读性
    % \resizebox{\textwidth}{!}{%
    \small
    \begin{tabular}{l cccc c cccc}
        \toprule
        & \multicolumn{4}{c}{\textbf{(a) Transmission Ranges}} & & \multicolumn{4}{c}{\textbf{(b) Sharing Ranges}} \\
        % & \multicolumn{3}{c}{\textit{Levels: Local, Home, Internet}} & & \multicolumn{3}{c}{\textit{Levels: Local, Company, Third-Party Company}} \\
        \cmidrule(lr){2-5} \cmidrule(lr){7-10}
        \textbf{Category} & \textbf{$\chi^2(2)$} & \textbf{$p$} & \textbf{$W$} & \textbf{Agreement} & & \textbf{$\chi^2(2)$} & \textbf{$p$} & \textbf{$W$} & \textbf{Agreement} \\
        \midrule
        Multimedia  & 643 & $<.001^{***}$ & .780 & \textbf{Very Strong} & & 673 & $<.001^{***}$ & .817 & \textbf{Very Strong} \\
        Health      & 630 & $<.001^{***}$ & .760 & \textbf{Very Strong} & & 398 & $<.001^{***}$ & .483 & Moderate\\
        Contact     & 612 & $<.001^{***}$ & .740 & \textbf{Very Strong} & & 282 & $<.001^{***}$ & .342 & Moderate \\
        Consumption & 604 & $<.001^{***}$ & .730 & \textbf{Very Strong} & & 688 & $<.001^{***}$ & .835 & \textbf{Very Strong} \\
        Search      & 666 & $<.001^{***}$ & .810 & \textbf{Very Strong} & & 657 & $<.001^{***}$ & .797 & \textbf{Very Strong} \\
        \midrule
        \textbf{Overall} & \textbf{132} & \textbf{$<.001^{***}$} & \textbf{.160} & Weak & & \textbf{126} & \textbf{$<.001^{***}$} & \textbf{.153} & Weak \\
        \bottomrule
        \multicolumn{10}{l}{\footnotesize{*: $p < .05$, **: $p < .01$, ***: $p < .001$.}} \\
        % \multicolumn{8}{l}{\footnotesize{\textit{Post-hoc tests confirm a uniform risk pattern:}}} \\
        % \multicolumn{8}{l}{\footnotesize{\textbf{(a)} Local $\approx$ Home $\ll$ Internet \quad \textbf{(b)} Local $\approx$ Pro $\ll$ Third-Party}} \\
        % \multicolumn{8}{l}{\footnotesize{(`$\approx$' indicates $p > .05$; `$\ll$' indicates $p < .001$.)}}
    \end{tabular}%
    % }
\end{table*}

Qualitative analysis reveals that this boundary carries rich social meaning. \textbf{Participants consistently viewed the home as a psychological private space, viewing data leaving this boundary as a violation of personal space.} P39 articulated this sense of intrusion directly, \textit{``Sending my data out feels different ... like letting something unknown into my private space ...''} P28 echoed this, emphasizing the emotional dimension: \textit{``It just feels fundamentally wrong for private conversations, things said in the home, to end up on some random server somewhere...''} This highlights home as a personal and relational private space. 

\textbf{The boundary meant a zone of perceived control.} Within the home network, users felt a strong sense of agency over their data. Crossing the boundary signified entering an uncontrolled, unpredictable domain. As P3 stated, \textit{``Inside my home, I feel like I have some control. Once it's out? it's total out of my hands.''} This loss of control is directly linked to the introduction of unknown social actors, which P13 identified as a proliferation of  \textit{``more individuals who could potentially acquire the data''}.

\textbf{Consequently, the home boundary functioned as a heuristic for risk assessment.} P4's described using this boundary as a primary filter, stating that if data leaves the home, their \textit{``guard goes way up immediately''}. This shift in perception was not merely incremental but categorical. P30 emphasized that external transmission represents \textit{``a complete different category of risk''}, suggesting that users construct distinct mental rules to govern the uncertainty of the external world.

\textbf{Quantitative analysis further corroborates that the risk escalation associated with the home boundary remains consistent across divers functional data categories.} As detailed in Table~\ref{tab:merged_stats}, the boundary effect overshadowed specific data content in shaping perceived privacy risk. Post-hoc Nemenyi analyses with Bonferroni corrections confirmed significant differences ($p < .05$ to $p < .001$) primarily when comparing in-home flows against those transmitted to broad internet. This validates the robustness of the home as a privacy boundary.
% This uniformity validates qualitative view that internet transmission triggers a universal shift in perceived privacy risks.
%indicates that the boundary effect overshadows specific data content in shaping risk perception, validating users' view that internet transmission triggers a fundamental shift in privacy assessment.

% heuristic (\textit{``My first question is always: Does it leave [the home] ... ? It yes, my guard goes way up immediately ...''}) shows how location acts as a primary rule for boundary regulation, characteristic of managing personal space. The perceived shift was not merely incremental but categorical (P30: \textit{``... a completely different category of risk ...''}), signifying the construction of rules governing the uncertain world.

% Remarkably, this pattern of risk escalation at the network boundary held consistently across diverse functional data categories (see Table~\ref{tab:merged_stats}). Post-hoc analyses confirmed significant differences ($p < .05$ to $p < .001$) primarily when comparing within-company flows to those shared to third-party companies. This uniformity indicates the boundary effect overshadows data specifics in shaping users' risk perception. Users confirmed this, with P4 asking \textit{``Does it leave...? If yes, my guard goes way up immediately...''}, and P30 calling internet transmission \textit{``a completely different category of risk''}. These perspectives shows the existence of the boundary as a heuristic, simplifying risk assessment into location-based zones before content scrutiny. 

\subsection{The Provider to Third-party Threshold Acts as a Critical Sharing Boundary} \label{result:sharing}

\textbf{Data sharing exhibits a critical privacy boundary when moving from the primary service provider to third-party providers.} Risk perception remained relatively low when data was kept locally (M=4.59) or shared only within the primary service provider (M=4.66). Risk escalated significantly when data was shared with third-party companies (M=4.85) (see Figure~\ref{fig:transmission_sharing}). A Friedman test confirmed this increase (see Table~\ref{tab:merged_stats}), with post-hoc Nemenyi analyses revealing significant differences between third-party sharing and the other two conditions ($p < .05$). Similarly, the effect was significant across data categories (all $p < .001$ for main effect, and for post-hoc comparisons company vs. third-party company $p < .05$). 
% This boundary crossing triggered strong negative reactions rooted in social norms of trust, reciprocity and accountability. 

\textbf{Qualitatively, this distinction in risk is driven by users' categorization of recipients based on the perceived relationship}, who viewed third parties as fundamentally different from the primary provider. While participants maintained a wary but accepted relationship with the primary provider, they viewed third parties as fundamentally different. P8 labeled these external companies as \textit{``strangers''}, noting that they represent a \textit{``totally different category''}. This distinction establishes a clear social boundary that defines the rules of data permeability.

\textbf{Participants frequently framed third-party sharing as a violation of the agreement for them and the smart assistant (i.e., data co-owner).} Users described this distribution of data as a breach of social contract. P35 likened the behavior to the assistant \textit{``gossip[ping] with advertisers,''} while P38 stated that such actions \textit{``break the expected confidence.''} This perceived breach was closely linked a sense of betrayal. P10 acknowledged the functional necessity of data access but questioned the legitimacy of the provider \textit{``selling it off to who-knows-who.''} P23 captured this sense by comparing the scenario to \textit{``hiring a bodyguard who then invites strangers into your house,''} highlighting the violation of relational expectations.

\textbf{Transferring data to third parties also signaled a drastic loss of control and accountability.} Participants expressed concern about the untraceable propagation of their information, with P40 fearing the data \textit{``branches infinitely''} and P12 describing the diffusion as \textit{``dripping ink into water.''} This lack of visibility made it difficult for users to assign responsibility. P6 noted that unlike the primary provider, there is no clear point of contact \textit{``if something goes wrong''} with unknown entities. Furthermore, users expressed discomfort with the commodification of their personal context. P12 found it unsettling that companies were \textit{``packaging up bits of my private life''} and treating intimate data as a tradeable asset.

\section{RQ2: The Nuanced Interaction of Transmission and Sharing Ranges with Contextual Factors}\label{sec:rq2}

PBT explains that individuals manage privacy boundaries through rules derived considering: (1) attributes of the private information (Section~\ref{sec:rq2_attribute}), (2) perceived risk-benefit assessments (Section~\ref{sec:rq2_risk}), (3) demographic factors (Section~\ref{sec:rq2_demo}) and (4) technical measures affecting boundary control (Section~\ref{sec:rq2_tech})~\cite{petronio1991communication,petronio2002boundaries}. Our findings systematically investigate how perceived risk with SPAs varies across these core factors.

\subsection{Privacy Boundary Varied Across Attributes of the Private Information}~\label{sec:rq2_attribute}

Aligning with PBT, our findings reveal that information attributes related to personal sensitivity and interpersonal connection influence perceived risk. This offers a lens for managing the unique risks posed by socially interactive agents~\cite{oh2024better}. We detail two primary findings: extend privacy boundaries to incorporate inferred data, and boundary rule application exhibits significant context-dependency.

\subsubsection{Extend Privacy Boundaries to Incorporate Inferred Data}

\textbf{Participants extend privacy boundaries beyond explicitly provided data to encompass information inferred by the system.} Data inferred about users' habits, home environment (e.g., during voice games) or personal routines (e.g., derived from schedule planning for recommendations) were perceived as highly sensitive, particularly when such data was subject to transmission outside the home (M=5.00 for home environment, M=4.94 for schedule planning) or shared with third parties (M=4.99 and M=4.91, respectively), both significantly higher than when transmitted within the home or with the service provider ($p < .001$ via Friedman tests with post-hoc Nemenyi comparisons). Participants viewed such inferences as intrinsically personal matters that they did not wish others to know (P20). Consequently, they expressed strong aversion to potential negative outcomes, such as the creation of detailed profiles that disgusted their experience of the system (P39). This suggests a subjective boundary definition aimed to protect the meaning derived from data interactions. It reflects a sense of ownership over inferred insights, serving as a defense against unwanted data linkages and the anticipated turbulence resulting from the misuse of profiles exposing their states.

% (\textit{``also personal things I did not want others to know,''} (P20)) and expressed strong aversion to potential negative consequences like unwanted profiling (\textit{``profiling which disgusts my experience''} (P39)). This shows a subjective boundary definition aimed at protecting the meaning derivable from data interactions. It reflects perceived ownership over inferred insights and serves as a defense against unwanted boundary linkages and the anticipated ``turbulence'' resulting from the misuse of inferred profiles about their core self or internal state. 

\subsubsection{Boundary Rule Application Exhibits Significant Context-dependency}

\textbf{The application of boundary rules is influenced by the inherent sensitivity of specific data types, resulting in significantly different baseline risks.} Friedman analyses and Nemenyi post-hoc analyses revealed statistically significant variations in perceived privacy risks across communication, entertainment, health, consumption and daily task data types ($\chi^2_4 = 25.349$, $p < .001$, $W = 0.015$). Communication-related data evoked the highest mean risk rating (M=4.89), significantly exceeding other categories ($p < .05$), whereas daily task data yielded the lowest (M=4.30, $p < .05$). Health (M=4.59), consumption (M=4.64), and entertainment (M=4.66) exhibited intermediate risk levels ($p > .05$ compared with others). This differentiation supports PBT's proposition that information attributes influence users' sense of ownership and control. As P39 articulated: \textit{``Shopping may reveal personal preferences ... but it is relatively low-risk compared to taking a taxi, which exposes one's travel route. Health data, however, is far more sensitive.''}

\textbf{Sensitivity is further modulated by the functional necessity of boundary crossings, distinguishing acceptable service exchanges from intrusive violations.} The permeability of established boundaries, such as the `home', demonstrates context-dependency. Transmitting sensitive historical information (e.g., smart Q\&A, $\Delta=0.47$, $p < .001$) or personal accounts (e.g., map viewing, $\Delta=0.56$, $p < .001$) outside the home significantly increased perceived privacy risk (P14, P8). Conversely, functionally essential boundary crossings, such as sharing \textit{location information} for ride-hailing ($\Delta=0.28$, $p < .05$), were less concerning, and considered a necessary \textit{``agreed-upon exchange for that specific service''} (P18), indicating successful boundary coordination in that context. Data flows deemed inappropriate for the context generated high perceived privacy risk. For example, P34 noted that \textit{``It doesn't make sense for my smart lightbulb data to end up with an insurance company, even if anonymized.''} However, functionally logical flows were accepted when the usage aligned with the app's purpose (P34). Even when initial sharing is deemed acceptable, concerns often shift towards subsequent usage, specifically managing the linkage regarding who received the data (P19), characterizing the privacy management as an ongoing negotiation rather than a static agreement (P19). 
% (e.g., \textit{``it doesn't make sense for my smart lightbulb data to end up with an insurance company, even if anonymized''}) generate high risk, whereas functionally necessary flows (e.g., \textit{``it makes sense for the map app to use my location''}) are accepted (P34). Even when initial sharing is deemed acceptable, concerns often shift towards subsequent data linkage and usage management (\textit{``managing the linkage: Who else gets this data? How it is used later?''}), representing an \textit{``ongoing negotiation''} (P19).

\begin{figure*}[!htbp]
    \includegraphics[width=0.75\textwidth]{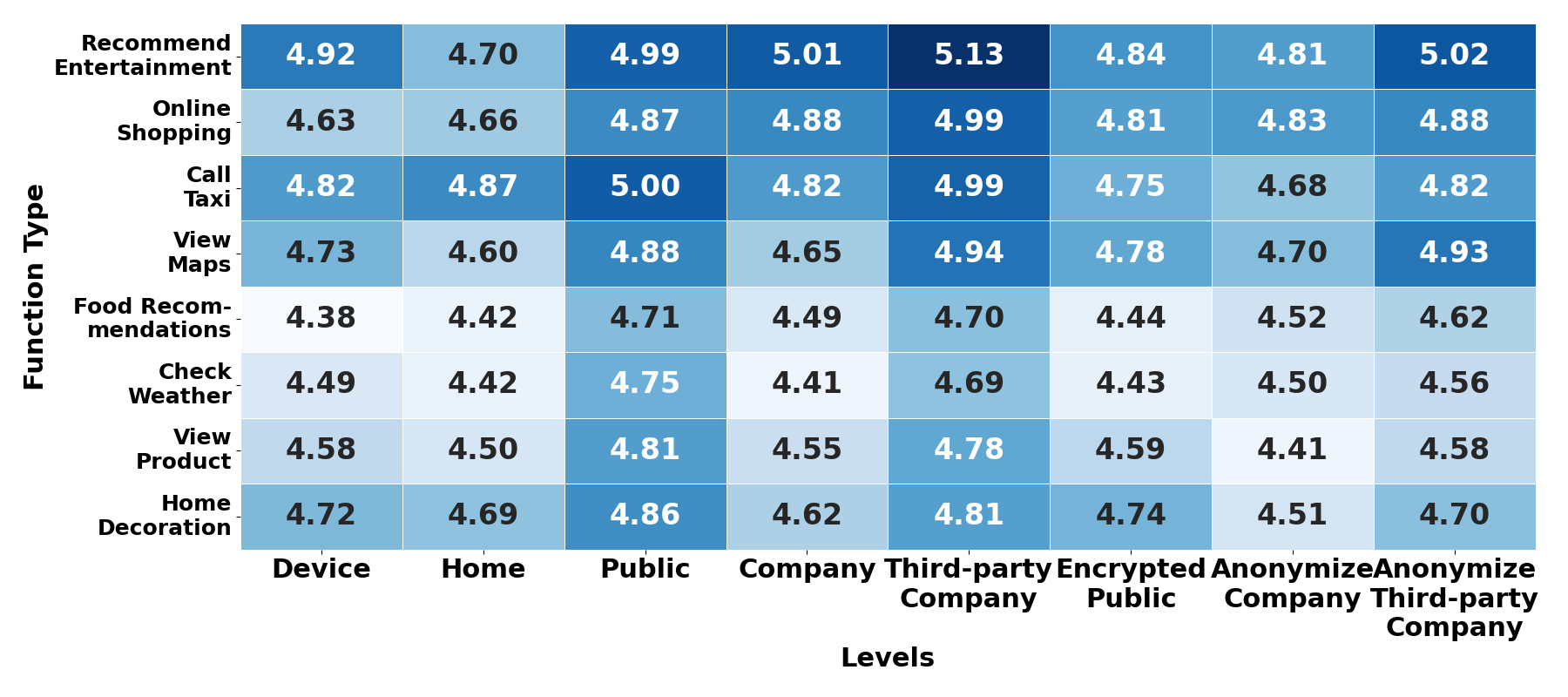}
    \caption{Perceived privacy risks of the target location information for different functions.}
    \label{fig:target_location}
\end{figure*}

\textbf{Finally, risk perception is complicated by the interaction between data type and function, confirming that neither factor operates in isolation.} Within a single function like online shopping, the context dictated higher risk for sharing inherently sensitive \textit{payment information} (M=5.28) compared to less sensitive \textit{personal preferences} (M=4.69, $p < .001$ compared to payment information). Conversely, the perceived privacy risk associated with the exact same data type, \textit{target location}, varied significantly by the function it served: it was perceived as with relatively low risk for food recommendations (M=4.65) but with significantly higher risk for online shopping (M=4.89, $p < .01$ compared to food recommendations) or viewing maps (M=4.90, $p < .01$ compared to food recommendations) (see Figure~\ref{fig:target_location}). Indeed, location data's perceived privacy risks spanned a wide spectrum, particularly when transmitted across the broad internet (M=4.71-5.00) or shared with third-party companies (M=4.69-5.13). This highly context-specific evaluation aligns with PBT's boundary coordination concept. Participants indicated that data flows conforming to their expectations exemplify successful coordination. P6 noted that such flows are \textit{``perceived as necessary,''} validating the exchange. However, this coordination is fragile, as concerns regarding potential secondary use often persist even after initial access is granted (P17). Conversely, deviations from purpose limitations were viewed as clear boundary coordination failures (P15). P10 explained that these deviations serve to \textit{``heighten risk perception, signal boundary turbulence, and erode trust''}, transforming a functional interaction into a privacy violation. Therefore, effective boundary management requires continuous, context-aware rule adjustment by the user, a process P10 summarized as \textit{``a constant balancing act''} between utility and exposure.
% Participants indicated that expected data flows are \textit{``perceived as necessary exemplify successful coordination''} (P6), although they noted that \textit{``concerns about potential secondary uses may persist even when initial access is granted''} (P17). Conversely, deviations from purpose limitations  \textit{``represent boundary coordination failures''} (P15), which consequently serve to \textit{``heighten risk perception, signal boundary turbulence, and erode trust''} (P10). Consequently, effective boundary management requires continuous, context-aware evaluation and rule adjustment by the user, a process described by P10 as \textit{``a constant balancing act''}.

\subsection{Perceived Risk-benefit Assessments Impact Risk Perception}~\label{sec:rq2_risk}

\begin{figure*}[!htbp]
    \subfloat[]{
        \includegraphics[width=0.5\textwidth]{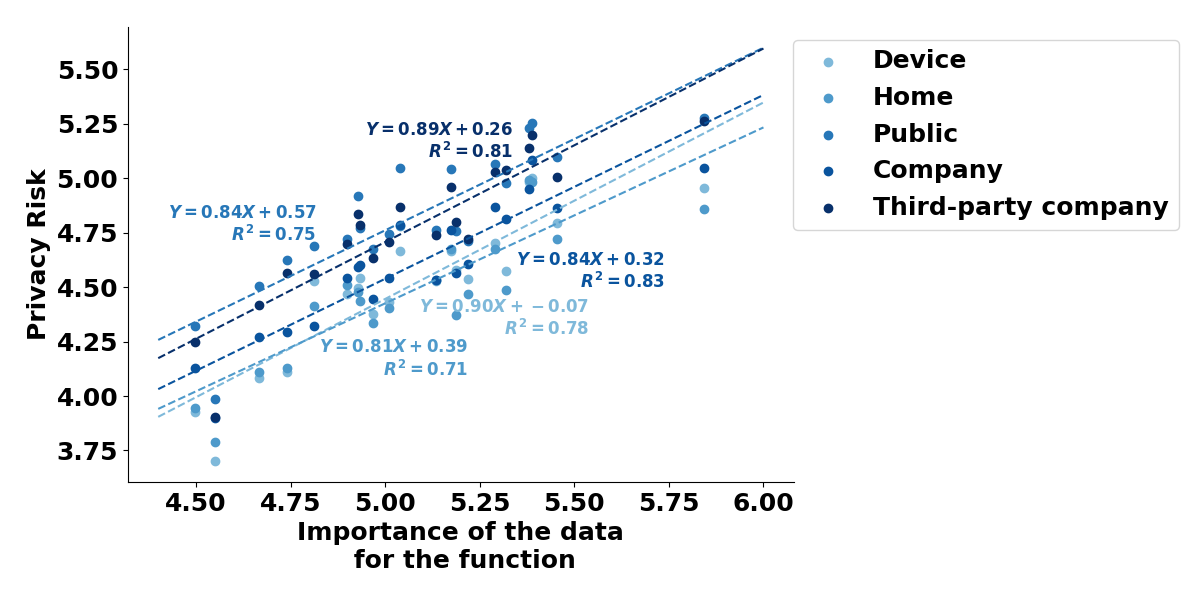}
        \label{fig:info_importance}
    }
    \subfloat[]{
        \includegraphics[width=0.5\textwidth]{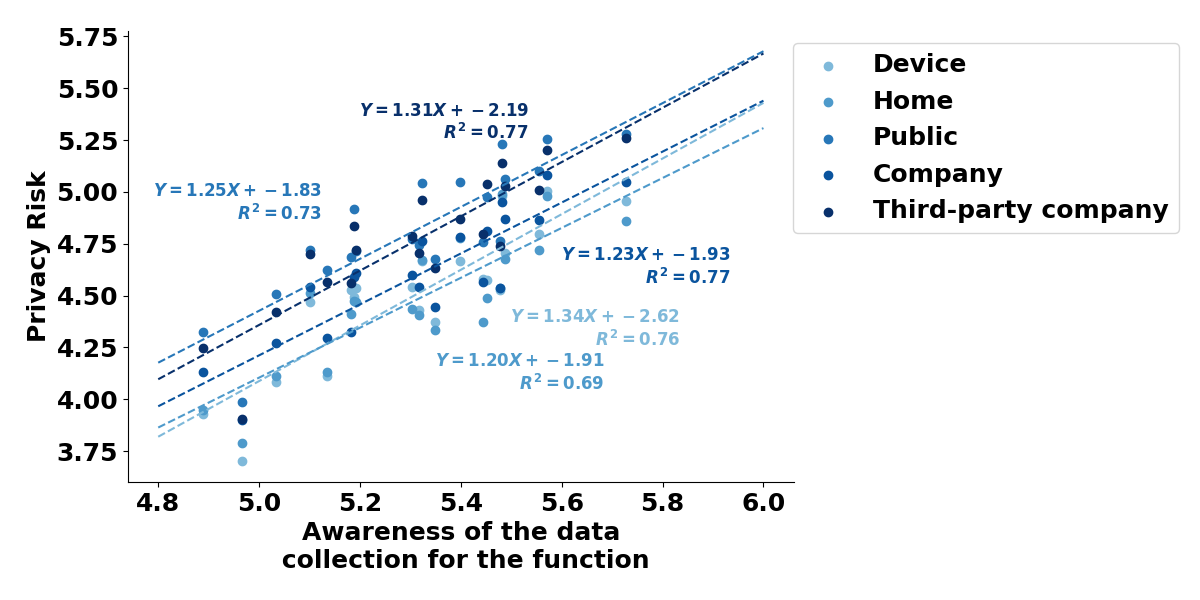}
        \label{fig:info_knowledge}
    }
    \caption{(a) Correlation between the importance of the data collected and privacy risk, (b) Correlation between the awareness of the data collected and privacy risk. The analysis was performed on the original 7-point Likert scale ratings.}
\label{fig:info}
\end{figure*}

Using correlation analyses, we found strong positive correlations between \textit{perceived privacy risk} and both the \textit{perceived importance of the data} and \textit{the awareness of its collection} (both $R^2 > 0.7$, Figures~\ref{fig:info_importance},~\ref{fig:info_knowledge}). Interpreting these findings through the lens of PBT~\cite{petronio1991communication,petronio2002boundaries} allows us to understand privacy management as a dynamic process of setting and adjusting privacy calculus rules according to the perceived value of data. 
% This view offers distinct insights to understand the nuances of user-SPA interactions.

\subsubsection{Risk-benefit assessments and boundary regulation}

\textbf{The perceived importance of data influences the privacy calculus, often necessitating boundary adjustments (see Figure~\ref{fig:info_importance})}. Consistent with PBT, users recognized a tension where data critical for functionality is simultaneously perceived as intrinsically sensitive, requiring an evaluation before crossing boundaries. P10 articulated this trade-off, noting that when functionality co-occurs with high perceived privacy risks, it demands a deliberate \textit{``calculus process''} regarding whether to disclose the information. This demonstrates how the risk-benefit assessment is important in how users negotiate privacy boundaries.
% sers articulate this linkage, recognizing that functionally critical data often feels intrinsically sensitive, where users adjust which boundaries the data was allowed to cross, as their calculus in response to this relative relationship between sensitivity and utility. As P10 described, \textit{``If the specific functionality is crucial, usually it's more private for me. However, I needed more calculus process whether to disclose the information to use the corresponding function.''} When high functional importance converges with high perceived sensitivity, users face significant internal conflict in setting different boundaries. 
% For exampled, P7's described the \textit{``constant internal fight: open the gate for the benefit, or keep it locked tight for safety?''} 

\subsubsection{Awareness and potential boundary turbulence}

\textbf{Similarly, user awareness of data collection practices exhibits a strong positive correlation with perceived privacy risk (Figure~\ref{fig:info_knowledge})}. A lack of awareness can lead users to underestimate or entirely overlook risks, as P6 admitted: \textit{``I realized that I never considered the question of `possibly needing to view my order history' previously. Because I had no knowledge about it, I didn't even realize there could be some privacy risk.''}

\textbf{Conversely, increased awareness heightens risk perception, particularly when revealed data practices violate contextual expectations, aligning with~\cite{li2021valuing, vitak2018privacy, schaub2015design}.} From a PBT perspective, when users discover data flows that contradict their established rules, such as a voice game accessing home environment data, it trigger boundary turbulence. P12 described such unexpected requests as \textit{``unreasonable''}, leading directly to increased perceived privacy risk. This indicates that transparency can paradoxically increase anxiety when it exposes discrepancies between user expectations and system behavior.

\textbf{Beyond static assessment, awareness may leads users to expend their privacy boundary.} Learning about previously obscure data flows can expand a user's conception of what constitutes private information, effectively enlarging their ``privacy zone''. P5 described this cognitive shift, noting that realizing smart light usage could reveal sleep patterns forced them to include energy data in their private boundary, making them feel \textit{``more vulnerable overall''}. 
% Furthermore, PBT illuminates awareness not merely as static input for risk calculation, but as a catalyst for dynamic boundary (re)definition. Learning about previously unknown data flows can expand the user's conception of their own private sphere. P5 described this process: \textit{``Learning that my smart light usage could reveal my sleep patterns made me suddenly realize that energy data is part of my private boundary now ... My `privacy zone' keeps getting bigger as I learn more ..., which makes me feel more vulnerable overall.''}

\subsection{Demographic Factors have Limited Influence on Perceived Privacy Risk}~\label{sec:rq2_demo}

Our analysis reveals that core demographic factors, such as age, gender and education, exert limited main influence on perceived privacy risks associated with shifts in data transmission or sharing ranges. Quantitative analysis using Kruskal-Wallis H tests showed no significant main effects for age (see Table~\ref{tab:demographics_anova}). This provides compelling evidence for the robustness of privacy boundaries. It suggests that core distinctions central to perceived privacy risk, such as data remaining `inside' versus moving `outside' the home, or data shared with a `known' primary provider versus `unknown' third parties are deeply ingrained and operate consistently across demographic groups. Participants articulated these distinctions as self-evident truths. As P8 remarked: \textit{``it's just ... obvious, isn't it? Stuff inside my house should stay inside... data going to the company I bought the device from is one thing. data going to random companies is completely different. It just feels fundamentally different.''}

\begin{table}[htbp]
\centering
\caption{Effects of demographics on perceived risk. Effect sizes ($\epsilon^2$) are classified as Negligible ($<.01$), Small ($<.06$), Moderate ($<.14$), or Large ($\ge.14$).}
\label{tab:demographics_anova}
\small
\begin{tabular}{l ccc c ccc}
  \toprule
  & \multicolumn{3}{c}{\textbf{Transmission Ranges}} & & \multicolumn{3}{c}{\textbf{Sharing Ranges}} \\
  \cmidrule(lr){2-4} \cmidrule(lr){6-8}
  \textbf{Factor} & $\epsilon^2$ & \textbf{Effect Size} & $p$ & & $\epsilon^2$ & \textbf{Effect Size} & $p$ \\
  \midrule
  Age        & .016 & Small & .24 & & .007 & Negligible & .70 \\ 
  Gender     & .003 & Negligible & .59 & & .003 & Negligible & .54 \\
  Education  & .003 & Negligible & .85 & & .009 & Negligible & .47 \\
  \bottomrule
  % \multicolumn{8}{l}{\footnotesize{\textit{Note: Kruskal-Wallis H tests indicated no significant differences across groups.}}}
\end{tabular}
\end{table}

\textbf{These privacy boundaries exhibit remarkable stability, contrasting with prior work that often emphasizes the heterogeneity of privacy preferences~\cite{asthana2024know}.} While PBT accounts for contextually negotiated rules, our results suggest that privacy boundaries serve as a common heuristic for privacy risk assessment. P17 described this as \textit{``ground zero''} for privacy evaluations, noting that \textit{``nobody argues that data staying in your house is different from data going out. That's just ... ground zero. We will start from here.''} 
% Therefore, while demographics might influence nuanced, context-specific privacy rule negotiations or risk tolerances, their roles appear minimal in shaping perceptions of these fundamental boundary transformations.

\subsection{Technical Measures Fail to Mitigate Boundary Permeability}~\label{sec:rq2_tech}

We employed Friedman tests, and post-hoc Nemenyi tests with Bonferroni comparisons to evaluate how technical safeguards modulate risk perception across transmission and sharing ranges. Our analysis reveals that technical safeguards offer limited risk mitigation, where encryption's effectiveness reduces with wide transmission, anonymization fails to mitigate trust for third-party sharing, and data attributes moderate anonymization's effectiveness, highlighting boundary coordination failures.

\subsubsection{Encryption's Effectiveness Reduces with Wide Transmission} 

\textbf{Explicitly mentioning encryption significantly mitigates perceived risks associated with data transmission but fails to address concerns regarding subsequent data access.} Quantitative results show that encryption reduced the perceived risk of Internet transmission to levels comparable to local transmission (see Table~\ref{tab:mechanism_summary_combined}), suggesting it effectively addresses the boundary breach associated with spatial movement. However, participants argued that encryption protects data only during transit without offering control over the recipient's actions once the data is decrypted. For example, P13 noted the limitation that encryption addresses the \textit{means} of transport rather than the \textit{terms} of access. Consequently, to be effective, encryption should be paired with visible, verifiable usage controls that persist after the data reaches its transmission destination.

\subsubsection{Anonymization Fails to Mitigate Distrust For Third-Party Sharing}

\textbf{Anonymization demonstrates a context-dependent effect, proving effective within established relationships but losing potency with third parties.} When data is shared within the primary service provider, anonymization significantly lowers perceived risk to levels statistically indistinguishable from local storage (see Table~\ref{tab:mechanism_summary_combined}). In this context, users accept anonymization as a valid boundary control mechanism because it operates within an existing relationship of partial trust. 

\textbf{Crucially, this reassurance collapses when data is shared with third parties.} Although anonymization still a statistically significant risk reduction compared to non-anonymized sharing, the absolute risk levels remain significantly higher than saving the data on-device. This failure stems from a breakdown in boundary coordination~\cite{petronio2002boundaries}. Users view third-party anonymization as a unilateral declaration by the provider rather than a mutually verified rule. Therefore, they expressed skepticism regarding the motives of unknown entities and the permanence of the safeguard. For example, P33 noted that \textit{``third-party goals aren't necessarily aligned with user interests''}, and \textit{``anonymous today might not be anonymous tomorrow.''}

\textbf{Furthermore, the inherent opacity of these measures exacerbates distrust.} Users described anonymization as a ``black box'' promise lacking assurance of compliance. P14 argued that without knowing if the recipient agrees to the rules, \textit{``anonymized doesn't mean much''}. This indicates that current anonymization practices fail not necessarily due to technical inadequacy, but because they lack the transparency and enforceability required to establish a trusted boundary linkage with unknown third-party parties.

\begin{figure}
    \includegraphics[width=0.5\textwidth]{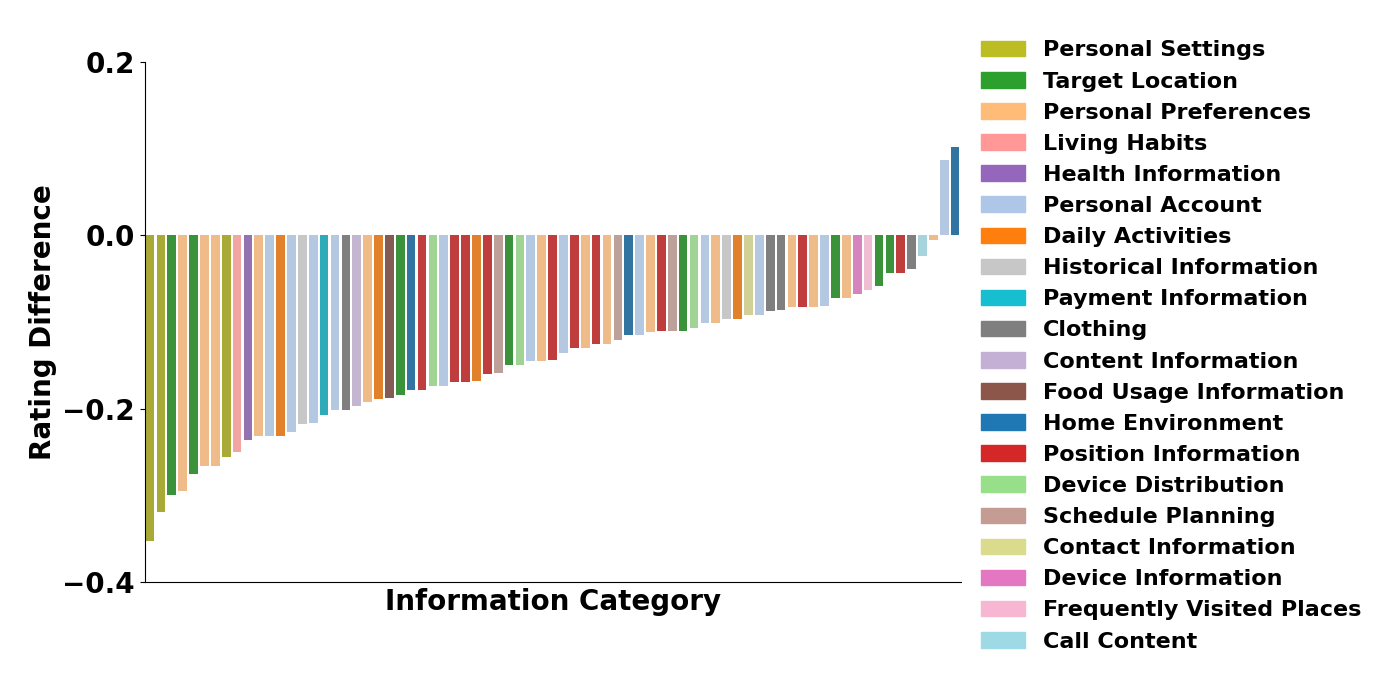}
    \caption{Difference of risk ratings for data shared with third-party companies between two conditions: anonymized versus non-anonymized. Different colors represent distinct information categories. The analysis was performed on the original 7-point Likert scale ratings.}
    \label{fig:anonymize_difference}
\end{figure}

\begin{table}[h!]
    \centering
    \caption{Effectiveness of Encryption and Anonymization across five data categories. We test if the mechanism significantly lowers risk compared to the not stated condition, and if the protected condition is statistically indistinguishable from the Home/Local condition.}
    \label{tab:mechanism_summary_combined}
    \resizebox{0.5\textwidth}{!}{%
    \small
    \begin{tabular}{l ccccc}
        \toprule
        & \multicolumn{5}{c}{\textbf{Significance of Difference ($p$-value)}} \\
        \cmidrule(lr){2-6}
        \textbf{Comparison} & \textbf{Multimedia} & \textbf{Health} & \textbf{Contact} & \textbf{Consumption} & \textbf{Search} \\
        \midrule
        \multicolumn{6}{l}{\textit{\textbf{Encryption}}} \\
        \quad Encrypted vs. Internet & $\mathbf{<.001}^{***}$ & $\mathbf{<.001}^{***}$ & $\mathbf{<.001}^{***}$ & $\mathbf{.002}^{**}$ & $\mathbf{<.001}^{***}$ \\
        % \quad \textit{Effect Size ($r_{rb}$)} & .294 & .390 & .367 & .265 & .548 \\
        % \addlinespace[0.2em]
        \quad Encrypted vs. Home & .633 (ns) & .974 (ns) & .498 (ns) & .551 (ns) & .647 (ns) \\
        
        \midrule
        \multicolumn{6}{l}{\textit{\textbf{Anonymization (Third-Party Sharing)}}} \\
        \quad Anon. vs. Third-party sharing & $\mathbf{.044}^{*}$ & $\mathbf{.029}^{*}$ & $.052$ & $\mathbf{<.001}^{***}$ & $\mathbf{.002}^{**}$ \\
        
        % \addlinespace[0.2em]
        \quad Anon. vs. Local & $\mathbf{.045}^{*}$ & .106 (ns) & $\mathbf{.043}^{*}$ & $\mathbf{.015}^{*}$ & $\mathbf{.006}^{**}$ \\

        \midrule
        \multicolumn{6}{l}{\textit{\textbf{Anonymization (Service Provider)}}} \\
        \quad Anon. vs. In-company & .285 & $\mathbf{.046}^{*}$ & .280 & $\mathbf{.002}^{**}$ & $\mathbf{.037}^{*}$ \\
        \quad Anon. vs. Local & .358 (ns) & .773 (ns) & .636 (ns) & .212 (ns) & .830 (ns) \\
        \bottomrule
        \multicolumn{6}{r}{\footnotesize{$^{*}p<.05, ^{**}p<.01, ^{***}p<.001$; ns: not significant.}}
    \end{tabular}%
    }%
\end{table}

\subsubsection{Data Attributes Moderate Anonymization's Effectiveness}

\textbf{Compounding this trust issue, the perceived effectiveness of anonymization is moderated by the specific attributes of the data (see Figure~\ref{fig:anonymize_difference})}. Anonymization offered minimal perceived risk reduction when sharing personal preferences (e.g., play sound games, $\Delta$=-0.01) or personal account data (e.g., health advice, $\Delta$=-0.05) with third parties. In contrast, it provided a substantial risk reduction for historical information (e.g., sending emails, $\Delta$=-0.22) or target location data (e.g., online shopping, $\Delta$=-0.15).

This variability suggests users evaluate anonymization's credibility based on the perceived inherent identifiability or sensitivity of the data itself. Some data types, like environmental data, are seen as intrinsically difficult or impossible to truly anonymize. For example, P4 argued that \textit{``I could hardly think out how to anonymize the home environment. Making generative replacements?''}. P16 also echoed that \textit{``...anonymizing data about my home's layout or sounds? That seems impossible. How can data not be identifiable? For that kind of data, `anonymization' feels like putting tape over a crack -- it doesn't really secure the boundary.''} This aligns with PBT, indicating that individuals assess boundary regulation mechanisms not abstractly, but in relation to the specific attributes of the private information, judging whether the safeguard credibly alters the boundary's properties for that information.

\section{Discussions and Implications}\label{sec:discussion}

\subsection{Evolving SPAs' Privacy Protection with Privacy Boundaries}\label{sec:protect}

To empower privacy protection with the evolving of SPAs, we contrast our findings with prior work across three dimensions: theoretical perspectives, cognitive heuristics, and technical implementations. 

\textbf{Theoretical perspectives: shift to boundary management.} The transition of SPAs to anthropomorphic social agents~\cite{luria2016designing,oh2024better} requires a shift in privacy protection strategies from regulating specific data flows, as guided by CI~\cite{nissenbaum2004privacy,apthorpe2018discovering,abdi2021privacy} toward managing structural boundaries. Our application of PBT novelly reveals the boundary-related rules as cognitive heuristics to manage risks. Our core finding--that risk perception escalates non-linearly upon crossing the `home' or `provider' boundaries (Figure~\ref{fig:transmission_sharing})--suggests that users view these not merely as data transmission events, but as important shifts in information co-ownership~\cite{petronio2002boundaries}. This pinpoints users' mental model, and extends beyond the general privacy concerns identified in prior work~\cite{lau2018alexa,zeng2017end}. Consequently, PBT and CI funtion complementarily to streamline privacy configurations: PBT governs the hierarchical management of major boundary crossings, while CI refines the appropriateness of specific data flows within those established boundary layers~\cite{abdi2021privacy}. %  by pinpointing users' exact mental model.

% This perspective complements approaches that often necessitate flows tailored to social contexts, like those informed by CI theory~\cite{nissenbaum2004privacy, abdi2021privacy}. PBT's emphasis on consistent boundary application across information types facilitates layered privacy configurations, potentially streamlining management compared to regulating individual flows. In practice, PBT can guide the hierarchical management triggered by crossing major user-defined boundaries, while CI can refine the appropriateness of data flows operating within an established boundary layer~\cite{abdi2021privacy}. 

Notably, we contributed to PBT by measuring boundaries in SPA context and examined the effects of data types, where the unique structures of in-home device connections~\cite{castelli2017happened}, the highly complicated third-party services~\cite{zeng2018user}, and a lack of extended communication~\cite{sannon2020just} were substantially different from the social contexts~\cite{sannon2020just,cummings2025self}. We also provided insights on users' judgments regarding different boundaries across different data types, which potentially provided a method to quantitatively analyze boundaries and build agents that adhere to human's boundary regulations in the future~\cite{guo2025not}.

\textbf{Cognitive heuristics: spatial and relational boundaries.} Despite prior work indicating that users often possess incomplete technical mental models of the cloud~\cite{abdi2019more,malkin2019privacy}, our results in Section~\ref{sec:rq1} challenged this assumption and found that users  demonstrated a consistent perception of the `home' boundary across diverse data categories. Consequently, existing mitigation strategies such as jammers~\cite{chen2020wearable} or network filters~\cite{olade2019smart} should evolve from generic blocking tools to \textit{boundary-aware} interfaces. Protections should avoid obfuscating the data behind seamless cloud integration, thereby reinforcing user's agency and addressing the anxiety caused by the ``invisible'' nature of SPA data access~\cite{meng2021owning}. 
% This challenge the assumption that users are entirely oblivious to network transmission, but rather apply a binary ``in/out'' heuristic.

Extending discussions on privacy calculus~\cite{vimalkumar2021okay,cho2020will,fan2024evaluating}, we revealed in Section~\ref{sec:rq1} that while users exhibit a willingness to negotiate boundaries with the primary provider, viewing them as a legitimate co-owner of data, this trust does not extend to third-party companies. This contradicts the notion that users generally tolerate broad data sharing for ecosystem benefits~\cite{manikonda2018s}. Instead, the introduction of third parties is perceived as a breach of the interaction and data sharing with the agent, echoing `boundary turbulence' in PBT~\cite{petronio2002boundaries}. This suggests that the current industry standard of bundling third-party consent with general terms could be improved. The third-party sharing could be decoupled from provider consent as a separate, high-friction boundary crossing that requires explicit renegotiation~\cite{malkin2022runtime}, rather than passive acceptance.

\textbf{Technical implementations: transform to boundary-adaptive architectures.} We highlight the limitations of technical mitigation when boundaries are violated. While anonymization reduces concerns during sharing (Section~\ref{sec:rq2_tech}), it fails to eliminate risk reception when data is shared with third parties. This adds nuance to the privacy paradox~\cite{kokolakis2017privacy} that users calculate risks based on recipient \textit{identity}, not just data identifiability. The skepticism toward anonymization in third-party contexts (Figure~\ref{fig:anonymize_difference}) challenges the sufficiency of technical solutions in the absence of institutional trust~\cite{liao2019understanding,zheng2018user}, calling for tangible protection methods~\cite{windl2023investigating}. 

Given these empirically validated boundaries, such as in Section~\ref{sec:rq2_attribute}, we propose hybrid, boundary-adaptive architectures. Rather than static local-only (e.g., Snips~\cite{coucke2018snips}, KIMYA~\cite{de2023hey}) or cloud-only models, systems should dynamically determine processing locations based on data sensitivity, tolerating cloud processing for functional data but restricting personal data (e.g., health) locally, as guided by Figure~\ref{fig:target_location}. 
% These empirically validated boundaries support a move towards hybrid, boundary-adaptive architectures. Unlike static local-only models (e.g., Snips~\cite{coucke2018snips}, KIMYA~\cite{de2023hey}) which may compromise utility, or cloud-only models which compromise privacy, we suggest a dynamic approach. Since risk perception varies significantly by data attribute (Figure~\ref{fig:target_location})--where users tolerate cloud processing for functional data (e.g., weather) but resist it for personal data (e.g., health)--systems should dynamically determine the processing location (Section~\ref{sec:rq2_attribute}).

\subsection{Cultural Effects on Privacy Boundaries}

Privacy boundary regulation is not a universal construct but is rooted in cultural norms~\cite{liu2018regulate,liu2015culturally}. Therefore, we interpret our findings through Hofstede's six-dimensional framework~\cite{hofstede2011dimensionalizing} to hypothesize how cultural norms may modulate boundary formulation. While the ``home'' boundary appears consistent, we posit that these boundaries' permeability are likely subject to cultural variations. We discuss these specific modulations below:

\textbf{First, \textit{Collectivist} nature of Chinese society~\cite{liu2018regulate,liu2015culturally} likely reinforces ``Home Network'' boundary effect we found in Section~\ref{result:transmission}.} As the family defines the primary privacy sphere, we hypothesized that users exhibit lower concern for intra-home transmission compared to individualist contexts where domestic privacy is more contested~\cite{abdi2019more,huang2020amazon}. Therefore, collectivism may culturally amplify the home's role as a cohesive sanctuary.
% In this context, the family unit, rather than the individual, often defines the primary privacy sphere. This explains why our participants viewed the home as a cohesive sanctuary, exhibiting lower concern for intra-home data transmission compared to individualist Western contexts, where intra-home privacy is often a fiercer ground~\cite{abdi2019more,huang2020amazon}. Thus, while the home boundary is universal, its permeability may be culturally amplified by collectivism. 

\textbf{Second, \textit{Uncertainty avoidance} may amplify the risk surge at the ``Third Parties'' boundary we found in Section~\ref{result:sharing}~\cite{hofstede2011dimensionalizing}}. A lower tolerance for unknown entities~\cite{liu2018regulate}, typically of high uncertainty avoidance culture like Chinese, likely frames third-party sharing as a transition into an uncontrolled domain. This contrasts with cultures of lower uncertainty avoidance, where risk gradients may appear more linear~\cite{abdi2021privacy}.
% Chinese culture typically scores high in uncertainty avoidance, correlating with a lower tolerance for ``unknown'' entities~\cite{liu2018regulate}. This trait likely amplifies the perceived danger of third-party sharing, viewing it not just as a data hand-off but as a transition into an uncontrolled domain. In cultures with lower uncertainty avoidance (e.g., the UK), users might perceived the risk gradient as more linear~\cite{abdi2021privacy}. 

\textbf{Third, \textit{Power distance} also influences privacy boundary management.} High power distance suggests users perceive providers as authoritative, increasing the permeability of provider-facing boundary as a legitimate exercise of authority~\cite{stanton2003information,he2025exploring,cecere2015perceived}. However, this difference does not extend to unknown third parties, thereby forming the boundary between the provider and third parties.
% In high power distance cultures like China, users often perceive the service providers as authoritative entities rather than peers. This perception increases the \textit{permeability} of the provider-facing boundary, as users accept data flows as a legitimate exercise of authority~\cite{stanton2003information,he2025exploring,cecere2015perceived}. Conversely, boundaries remain rigid against unknown third parties, explaining the risk gap between provider and third-party sharing.

\textbf{Fourth, \textit{Masculinity vs. femininity} may influence the trade-off between functionality and privacy.} 
Chinese emphasis on efficiency in \textit{Masculine} cultures~\cite{hofstede2011dimensionalizing} suggests distinct privacy calculus, where users may relax boundaries when transmission achieves specific utility (e.g., security efficiency)~\cite{he2025exploring}. % Chinese society's inclination towards the masculine dimension, characterized by a priority on achievement, efficiency, and material success~\cite{hofstede2011dimensionalizing}, may amplify the privacy calculation in boundary formulations. As evidenced by the variation in risk perception across functions in Figure~\ref{fig:target_location}, users willingly relax boundary restrictions when data transmission is important to achieving specific utility goals (e.g., efficiency in security)~\cite{he2025exploring}, viewing boundary regulation as a tool for trade-offs rather than concealment.

\textbf{Fifth, \textit{Long-term orientation} reframes boundary crossings as future-oriented investments.} Cultures valuing perseverance for future rewards such as China~\cite{hofstede2011dimensionalizing} may accept boundary permeability (e.g., data transmission for model training) as an investment to yield a personalized ecosystem, as evidenced in Section~\ref{sec:rq2_risk} where we found users' privacy calculus was modulated by the perceived importance of the data. Short-term oriented cultures may oppositely focus on immediate privacy preservation~\cite{abdi2021privacy}.
% Consequently, the perceived risk of current boundary violations may be mitigated by the pre-existing anticipation of long-term service optimization, contrasting with short-term cultures where preserving the immediate privacy takes precedence~\cite{abdi2021privacy}.

\textbf{Finally, \textit{Indulgence vs. restraint} may influence the boundaries around behavioral data.} Chinese culture of \textit{Restraint}~\cite{hofstede2011dimensionalizing,liu2018regulate} may heighten sensitivity toward behavioral data (Section~\ref{sec:rq2_attribute}). In this context, exposing private audio or behavioral patterns risks ``loss of face'', potentially amplifying boundaries against third-party access to avoid social judgment.
% As a society characterized by \textit{Restraint}, Chinese culture emphasizes strict social norms and the regulation of gratification~\cite{hofstede2011dimensionalizing,liu2018regulate}. This may amplify the finding in Section~\ref{sec:rq2_attribute} that users perceive heightened risks for behavioral data types, such as audio recordings. In Restraint cultures, exposing such private life goes beyond technical risks and may be viewed as causing ``loss of face''. Consequently, users may amplify boundaries against third-party access to behavioral data to avoid social judgment. 

\subsection{Social, Temporal and Demographic Variations of Privacy Boundaries}

While our empirical results in Section~\ref{sec:rq1} establish the foundations of privacy boundaries, we extend our discussions to their application within dynamic real-world contexts. By analyzing social dynamics, temporal evolution and demographic variance, we discuss how boundaries vary across contextual conditions, thereby providing the basis for a comprehensive understanding of privacy boundary regulation.

\textbf{First, the multi-user dimension of the domestic SPA ecosystem alters privacy boundary negotiations.} Unlike traditional social contexts characterized by direct communication~\cite{cummings2025self}, the presence of multiple occupants transforms privacy from an individual decision into a collective negotiation~\cite{alshehri2023exploring}, where social density and hierarchy modulate risk perception~\cite{he2025exploring}. The boundaries we found (Figure~\ref{sec:rq1}), in these shared spaces, are not static but are continuously reshaped by social interactions. Therefore, boundaries could consider the complex, shared nature of the smart home environment~\cite{he2025exploring} beyond individual-centric models.

\textbf{Second, privacy boundaries exhibit inherent temporal dynamics, shifting as user experience and context evolve.} They fluctuated based on factors such as habituation, frequency of interaction, and the changing value of services over time~\cite{chalhoub2021did}. Consequently, our work could also inform beyond identifying a specific boundary to a methodology for mapping the underlying relationship between user perceptions and boundary formation. By quantifying how these judgments respond to temporal variables, our framework could support the development of adaptive privacy agents capable of continuous alignment with users' evolving boundaries, ensuring sustained protection~\cite{zhang2025towards}. 

\textbf{Finally, demographic factors such as age and technical literacy likely influence how users enforce these boundaries.} Although our results show that the ``Home'' boundary is recognized across all demographics (Section~\ref{sec:rq2_demo}), specific groups may react to it differently. For older adults, the home often represents a private place. Consequently, data leaving the home may be perceived as a severer violation, suggesting a need for systems that default to keeping data local. Conversely, users with limited technical knowledge may struggle to perceive the ``Third-Party'' boundary (Section~\ref{result:sharing}). Our interviews indicate that these users sometimes misunderstand technical details, failing to distinguish between different service providers. Therefore, their boundaries may be more blurred, or scoped strictly within the home, to avoid data being transferred to any unknown entities.

\subsection{Design Implications}

To bridge the gap between users' mental boundaries (Section~\ref{sec:rq1}) and the opaque data flows of current commercial ecosystems (Section~\ref{sec:industry_results}), we propose actionable, boundary-aware designs across three dimensions: \textbf{system design}, \textbf{interaction design}, and \textbf{information visualization}.

\textbf{System Design: Boundary-based access control.}

In response to the sharp risk escalation observed at the ``Home'' and ``Provider'' perimeters (Figure~\ref{fig:transmission_sharing}), system architects should move beyond binary permissions to boundary-based control. 

$\bullet$ \textit{Example 1: Split the computing architecture.} For high-sensitivity data (e.g., audio recordings, health metrics, Section~\ref{sec:rq2_attribute}), systems should default to on-device processing, transmitting only abstract feature vectors rather than raw data to the cloud to respect the ``Home'' boundary.

$\bullet$ \textit{Example 2: Boundary-based sandboxing.} To address third-party distrust (Section~\ref{result:sharing}), systems can implement a ``Sandbox Mode'' that allow SPAs to execute third-party skills (e.g., a food delivery skill) while technically restricting that skill's access to the broader user profile or historical data~\cite{fraser2023user,chen2024empathy}. The sandbox could dynamically adjust boundary permeability based on interaction history, rather than on one-time static settings.

\textbf{Interaction Design: Communicating risks at privacy boundaries.}

Since technical safeguards may fail to mitigate users' perceived privacy risk in third-party contexts (Section~\ref{sec:rq2_tech}), interaction design should manage the boundary turbulence at these boundaries through explicit cues.

$\bullet$ \textit{Example 1: Explicit identity handoff warnings.} SPA should verbally announce data controller changes \textit{before} a transaction (e.g., \textit{``I am now connecting you to [third-party app] for this service ...''}), distinguishing the trusted provider from external entities.

$\bullet$ \textit{Example 2: Fine-grained boundary-based settings.} Privacy dashboards~\cite{thakkar2022would} should regulate data flows according to boundaries and enable users to set boundary-based access control rules, such as letting them to decide which third-party could access what information.

\textbf{Information Visualization: Mapping flows to privacy boundaries.}

To align with users' boundary-based mental models, visualizations should adopt topological data maps that mirrors user's privacy boundaries.

$\bullet$ \textit{Example 1: Spatial anchoring.} Privacy dashboards~\cite{thakkar2022would} should visualize data flow spatially, where data remaining on-device could be visually anchored to the home icon, while third-party transfers should be depicted as traversing a distinct, outer perimeter.

$\bullet$ \textit{Example 2: Semantic color-coding.} Unlike interfaces that color-code by data type (e.g., green for camera~\cite{habib2021toggles,zhang2024ghost}), designers should color-code by recipient boundary. A visual gradient should shift from Cool (Safe/Home) to warm (Caution/Third-party) as data crosses outward, visually illustrating the non-linear escalation of risk.

$\bullet$ \textit{Example 3: AR Boundaries.} As an exploratory direction, technologies like AR~\cite{prange2021priview,zhang2025evaluating} can concrete these invisible boundaries by overlaying virtual ``walls'' around the smart home hub, though technical implementation requires further user testing to validate efficacy.
% making the invisible perimeter of data transmission perceptible, 

\section{Conclusion}
As SPAs increasingly adopt anthropomorphic features and function as social agents within intimate spaces, understanding how users manage the resulting privacy boundaries becomes essential, necessitating frameworks like PBT. Grounded in PBT, this research investigated user risk perceptions in this context. A preliminary focus group study identified transmission range (managing spatial boundaries) and sharing range (managing relational boundaries) as dominant factors influencing perceived risk. Our main study (N=412 survey, N=40 interviews) then empirically examined the effect of privacy boundaries (RQ1): perceived risk escalates significantly when data crosses the `home network' perimeter or extends from the `provider' to `third-parties', heuristics consistently applied across data types and demographics. Addressing RQ2, we found risk perception is further modulated by information attributes reflecting relational sensitivity, risk-benefit assessments, technical measures and demographics. Crucially, technical safeguards like encryption and anonymization showed limited effectiveness, particularly for third-party sharing, undermined by fundamental user distrust and perceived coordination failures. Collectively, our findings highlight that boundary crossings are pivotal, revealing a complex, context-dependent boundary regulation process with implications for user-centric privacy design.

% \begin{acks}
%     This work was supported by the Natural Science Foundation of China (NSFC) under Grant No. 62472243.
% \end{acks}

\bibliographystyle{ACM-Reference-Format}
\bibliography{sample}

\appendix 

\section{Ethics Considerations}

We recognized potential ethical concerns in our work and rigorously adhered to ethical standards, including adherence to the Menlo \cite{bailey2012menlo} and Belmont \cite{beauchamp2008belmont} reports' principles, to address these concerns. Our studies, all approved by our Institutional Review Board (IRB), employed meticulous experimental design and data anonymization to ensure participant confidentiality. We engaged experts across fields for oversight and thoroughly informed participants about the studies' aims, allowing voluntary withdrawal at any point. Our efforts focused on respecting individuals, ensuring beneficence through giving future privacy enhancement implications, maintaining justice by equitably compensating participants. We respected all participants and allowed them to exit the experiment at any time if they felt uncomfortable. In all, our research aims to contribute to more effective PPMs, guided by ethical principles and legal compliance.

\section{Criteria for Classifying Contextual Factors}\label{sec:criteria}

1. The factor should be related to the smart home personal assistants' data collection and has correlation with human's cognition on the data collection. 

2. The factor should be directly or indirectly linked to the smart home personal assistant or its supporting company, such as data sharing range.

3. The factor should not be directly linked to the human users, such as personality, traits.

4. The factor should not be directly part of the whole of the content of the information, but should be related to the contextual specificity of the information.

% \section{Statistical Testing Results For The Main Study}\label{sec:statistical_testing}

% Table~\ref{table:comparison} showed the statistical testing results in the main study. 

% \begin{table}[ht]
% \centering
% \begin{tabular}{l|c|c}
% \hline
% \textbf{Category} & \textbf{Home} & \textbf{Device} \\ \hline
% Multimedia        & $p = .63$                               & $p = .74$                                \\ \hline
% Health            & $p = .97$                               & $p = .83$                                \\ \hline
% Contact           & $p = .50$                               & $p = .90$                                \\ \hline
% Consumption       & $p = .55$                               & $p = .84$                                \\ \hline
% Search            & $p = .65$                               & $p = .53$                                \\ \hline
% \end{tabular}
% \caption{Statistical results when comparing the data to the internet in an encrypted manner encrypted with transmitted within the home and within the device.}
% \label{table:comparison}
% \end{table}

\section{The Content of the Questionnaire}\label{sec:content_questionnaire}

The questionnaire contained the question for each function and each information. Specifically, we gave the description and an example of the function before the assessment of each information. Notably, the original questionnaire is in Chinese because the experiment is grounded in China. We translated the questionnaire to English without changing its meaning.

An example introduction is:

\textit{Please evaluate the category of multimedia and games. All the ratings towards the information and functions below belongs to multimedia and games. You need to rate three different scales: understanding, importance and privacy risk. Understanding refers to the degree of awareness regarding whether the corresponding function will collect the corresponding information in the given context. Importance refers to the assessment of whether the it is important and necessary for the corresponding function to collect the corresponding information. Privacy risk refers to the potential privacy risks and harm of an information collection type given all possible harmful use cases. For example, the information may be used by the company for illegal trading, overly user profiling, data breach, surveillance and tracking, etc.}

Notably, the harmful use cases were selected from the past literature (e.g., illegal trading \cite{warren2007stolen}, overly user profiling \cite{edu2020smart}, data breach \cite{edu2020smart,tabassum2019don} surveillance and tracking \cite{kroger2021data}), which identified different potential harms for personal data in the context of smart home and smart assistant.

An example introduction of the function is:

\textit{In this scenario, you use a smart voice assistant to receive recommendations for entertainment venues, activities, or movies. You interact with the assistant through spoken commands or questions. The assistant may processes the request, accesses relevant data, and provides personalized recommendations based on your preferences or past interactions. For example, you could ask for popular movie suggestions or city activities, and the assistant would offer curated options tailored to your interests.}

Then, an example item in the questionnaire is:

Recommending entertainment activities, venues, or movies may collect your historical information, as past entertainment choices are crucial for future recommendations. You should rate the following entries towards the collection of your historical information. The higher the score, the greater the understanding and importance, and the larger the privacy risk.

\begin{enumerate}   
    \item \textit{awareness: whether the participant is aware of the fact that the function would collect the corresponding information}
    \item \textit{importance: the importance of the information to complete the corresponding function}
    \item \textit{privacy risk of data saved and shared locally: the privacy risk when the data is saved on the local devices}
    \item \textit{privacy risk of data transmitted within the local network: the privacy risk when the data is transmitted within your home, but could potentially be transmitted to different devices in your home.}
    \item \textit{privacy risk of data transmitted across the entire network: the privacy risk when the data is transmitted to the entire broad network.}
    \item \textit{privacy risk of data shared internally within the company: the privacy risk when the data is shared within the company which provided the service of the SPA. For example, to be shared within Amazon for interacting with Amazon Alexa.}
    \item \textit{privacy risk of data shared with third-party companies: the privacy risk when the data is shared with other companies than the service provdiers. For example, to be shared with OpenAI for interacting with Google Assistant.}
    \item \textit{privacy risk of data encrypted and transmitted network-wide: the privacy risk when the data is transmitted with encryption.}
    \item \textit{privacy risk of data anonymized and shared internally: the privacy risk when the data is shared within the service provider with appropriate anonymization.}
    \item \textit{privacy risk of data anonymized and shared with third parties: the privacy risk when the data is shared with other companies than the service provider with appropriate anonymization.}
\end{enumerate}

\section{Usage of SPA and The Collected Information}\label{app:usage_information}

Tables~\ref{usage_spa} and ~\ref{tbl:information_collection} showed the usage of SPAs and the collected information. 

\section{Additional Results of Real-World Implementation Investigation}\label{app:real_world}

Tables~\ref{fig:merged_table} and ~\ref{real_world_info} showed the real world implementation status, where Table~\ref{fig:merged_table} showed the surveyed results of the transmission, sharing, encryption, and anonymization settings. Table~\ref{real_world_info} showed the information declared to be collected by the manufacturers.

\onecolumn
\begin{longtable}{p{50px}|p{125px}|p{200px}}
    % \centering
    \caption{Usage of SPA and the collected information. The third-party functions we finally selected were highlighted in red and orange, where we merged the items in orange of one single category to a single item. The citations were the origin we referred to. \label{usage_spa}}\\

    \toprule
        Category & Function & Information \\
    \midrule
    \endfirsthead

    \caption*{(Continued) Usage of SPA, the collected information, and their frequency. }\\
    \toprule
        Category & Function & Information \\
    \midrule
    \endhead

    \bottomrule
    \endfoot

        \multirow{7}{50px}{Personal basic functions} & Recording \cite{abdi2021privacy} & Content, History, Preference, Schedule, Voice \\ 
         & Check time \cite{lau2018alexa} & Location  \\
         & Check reminder \cite{abdi2021privacy} & History , Preference, Schedule \\
         & Set alarm \cite{lau2018alexa} & History, Preference, Schedule \\ 
         & Set Do not disturb & \\
         & Remind to take the key \cite{lau2018alexa} & Location \\
         & Remind to charge \cite{lau2018alexa} & Device information \\ \hline
        \multirow{7}{50px}{Multi-media Entertainment} & Accompany children/the elderly & Daily activity, Members, Personal information, Posture, Preference, Voice \\ 
         & \textcolor{red}{Voice game} \cite{lau2018alexa} & Device information, History, Home Layout, Preference, Voice\\
         & \textcolor{orange}{Music / Opera} \cite{abdi2021privacy,lau2018alexa} & History, Preference \\ 
         & \textcolor{red}{Recommend Entertainment} & History, Location , Preference, Schedule\\ 
         & \textcolor{orange}{Listen to radio} \cite{lau2018alexa} & Preference\\
         & Entertainment gesture recognition & Daily activity\\
         & Role in playing \cite{lau2018alexa} & Preference, Social habits\\ \hline
       \multirow{22}{50px}{Controlling Devices \cite{lau2018alexa}} & Light & Daily routine, Device information, Device layout, History, Home layout, Location, Lightness, Preference \\
         & Lock \cite{abdi2021privacy} & Device information, Device layout, Location, Personal information \\
         & Refrigerator \cite{rasch2013smart} & Device information, Preference\\
         & Air conditioner \cite{abdi2021privacy} & Amount of substance, Device information, Device layout, Home layout, Humidity, Location, Preference, Temperature\\ 
         & Sweeping machine & Amount of substance, Device information, History, Home layout, Preference\\
         & Hydroelectric switch & History, Usage information\\ 
         & Surveillance \cite{abdi2021privacy} & Daily activity, Daily routine, Preference\\
         & Automated watering & Humidity, Home layout, Plant condition, Plant layout, Plant preference, Temperature\\
         & Window & Device layout, Lightness, Schedule, Weather\\ 
         & Check safety of devices & Device information, Usage information\\
         & Rice cooker & Device information, Preference\\ 
         & TV \cite{rasch2013smart} & Device information, History, Preference\\
         & Monitor remaining stock of bathroom supplies & Preference \\ 
         & Pet feeding system & Pet information, Pet activities, Preference\\ 
         & Washing machine \cite{rasch2013smart} & Device information \\ 
         & Computer & Password\\
         & Router & Device information, Device layout \\ 
         & Ventilation fan & Device information, Usage information \\
         & Smart mattress & Daily routine \\ 
         & Humidifier & Preference \\
         & Clothes drying rack & \\
         & Speaker & History, Preference \\ \hline
        \multirow{6}{50px}{Health} & \textcolor{red}{Health recommendation} \cite{abdi2021privacy} & Account information, Daily routine, Health information, Preference\\
         & \textcolor{red}{Nutrition planning} & Food usage information, History, Location, Preference\\ 
         & Oxygen levels & Amount of substance\\ 
         & Sleep and health status \cite{abdi2021privacy} & Account information, Daily routine, Location, Preference\\ 
         & Heart rate and blood oxygen levels & Account information, Daily routine, Health information\\
         & Monitor water quality & Members, Preference\\ \hline
         \multirow{5}{50px}{Contact} & \textcolor{red}{Phone calling} \cite{abdi2021privacy} & Account information, Contact, Content, History, Voice\\
         & \textcolor{orange}{Send Message} & Account information, Contact, Content, History\\
         & \textcolor{orange}{Send E-mail} \cite{abdi2021privacy} & Account information, Contact, Content, History\\
         & \textcolor{red}{Reply WeChat} & Account information, Contact, Content, History\\
         & \textcolor{red}{Post updates} & Content, Password, Preference\\ \hline
         \multirow{2}{*}{Consumption} & \textcolor{red}{Shopping} \cite{abdi2021privacy,lau2018alexa} & Account information, Location, Payment information, Preference\\
         & \textcolor{red}{Call taxi} \cite{abdi2021privacy} & Account information, Location, Preference\\ \hline

         \multirow{15}{*}{Searching} & \textcolor{red}{Intelligent QA} & Account information, Content, History, Personal Information \\
         & \textcolor{red}{Travel routes, maps, and traffic conditions} & Account information, Common places, Location, Personal information, Schedule\\
         & \textcolor{red}{Wearing} & Clothes, Preference, Weather\\
         & \textcolor{red}{Food guides} & Account information, History, Location, Preference\\
         & \textcolor{red}{Weather} \cite{abdi2021privacy,lau2018alexa} & Location, Schedule, Temperature, Weather\\
         & \textcolor{red}{Price and discount} & Account information, Location, Preference\\
         & \textcolor{red}{Home decor advice} & Device information, Device layout, Home layout, Location\\
         & \textcolor{red}{News} \cite{lau2018alexa} & History, Preference\\
         & Textual and audiovisual materials & Preference\\
         & Translation & \\
         & \textcolor{red}{Documents, official papers, assignments} & Personal information, Preference\\
         & \textcolor{red}{Check order} & Account information, History, Preference\\
         & \textcolor{red}{Look for things} & Daily activity, Home layout\\
         & \textcolor{red}{Bills and accounts} \cite{abdi2021privacy} & Account information, Payment information\\
         & Remind how long since last coming home & Daily activity\\
\end{longtable}

\begin{table*}[!htbp]
    \centering
    \caption{Information collection of selected functions.}
    \label{tbl:information_collection}
    \begin{tabularx}{\textwidth}{p{80px}|p{120px}|X}
    \toprule
        Class & Function & Information \\ \midrule
        \multirow{3}{50px}{Multi-media and Entertainment} & Voice game \cite{lau2018alexa} & Device information, Home layout, History, Preference \\
         & Music / Opera / Broadcast \cite{abdi2021privacy,lau2018alexa} & History, Preference \\
         & Entertainment recommendation \cite{lau2018alexa} & History, Location, Preference, Schedule \\ \hline
        \multirow{2}{*}{Health} & Health recommendation \cite{abdi2021privacy} & Account information, Daily routine, Health information, Preference \\
         & Nutrition planning & Food usage information, History, Location, Preference \\ \hline
        \multirow{3}{*}{Contact} & Phone calling \cite{abdi2021privacy} & Account information, Contact, Content, History \\
         & Send Message or Wechat & Account information, Contact, Content, History \\
         & Post updates & Content, Password, Preference \\ \hline
        \multirow{2}{*}{Consumption} & Shopping \cite{abdi2021privacy,lau2018alexa} & Account information, Location, Payment information, Preference \\
         & Call taxi \cite{abdi2021privacy} & Account information, Location, Preference \\ \hline
        \multirow{12}{*}{Searching} & Intelligent QA & Account information, Content, History, Personal Information \\
         & Travel routes, maps, and traffic conditions & Account information, Common places, Location, Personal information, Schedule \\
         & Wearing & Clothes, Preference, Weather \\
         & Food guides & Account information, History, Location, Preference \\
         & Weather \cite{abdi2021privacy,lau2018alexa} & Location, Schedule, Temperature, Weather \\
         & Price and discount & Account information, Location, Preference \\
         & Home decor advice & Device information, Device layout, Home layout, Location \\
         & News \cite{lau2018alexa} & History, Preference \\
         & Documents, official papers, assignments & Personal information, Preference \\
         & Check order & Account information, History, Preference \\
         & Look for things & Daily activity, Home layout \\
         & Bills and accounts \cite{abdi2021privacy} & Account information, Payment information \\ \bottomrule
    \end{tabularx}
\end{table*}

\begin{table*}[!htbp]
\centering
\caption{The transmission, sharing, encryption and anonymization settings of different voice assistants with their corresponding manufacturers and models. ``Transmis.'' denotes and examines whether the information would be transmitted via wide area network. ``Share.'' denotes and examines whether the information would be shared to third-party companies. ``Encryp.'' denotes whether the information would be encrypted during transmission. ``Anony.'' denotes whether the information would be anonymized during sharing. ``Core'' denotes what agents or assistants in its core is supporting its functioning.}
\label{fig:merged_table}
\begin{tabularx}{\textwidth}{p{2.5cm}|p{3cm}|p{1.5cm}|p{1.5cm}|p{1.5cm}|p{1.5cm}|X}
\toprule
\textbf{Manufacturer} & \textbf{Model} & \textbf{Transmis.} & \textbf{Share.} & \textbf{Encryp.} & \textbf{Anony.} & \textbf{Core} \\ \midrule 
Amazon & Amazon Echo, Echo Dot, Echo Look, Echo Show, Echo Spot, Echo Plus, Echo Auto, Amazon Tap, Amazon Fire TV Cube & Part & All & All & Unclear & Alexa \\ \hline
Meta & Facebook Portal & Part & All & All & Unclear & Alexa \\ \hline
Sonos & Sonos One & Part & All & All & Unclear & Alexa \\ \hline
Invxia & Invxia Tribym & Part & All & All & Unclear & Alexa \\ \hline
Lenovo & Lenovo Smart Assistant & Part & All & All & Unclear & Alexa \\ \hline
Samsung & Samsung Galaxy Home & Part & All & Unclear & Unclear & Bixby \\ \hline
Microsoft & Invoke & Part & Part & All & Part & Cortana \\ \hline
Clova & Clova Wave, Clova Friends & Part & All & All & Unclear & Clova \\ \hline
Google & Google Nest Audio & Part & Part & All & Part & Google Assistant \\ \hline
Apple & Apple's Homepod, Apple's Homepod Mini & Part & No & All & Unclear & Siri \\ \hline
Mycroft & Mycroft Mark 1, Mycroft Mark 2 & Unclear & All & Unclear & Unclear & Mycroft \\ \hline
DingDong & DingDong & Unclear & Part & Unclear & All & DingDong \\ \hline
Baidu & Baidu Raven H, Baidu Xiaodu & Unclear & All & Unclear & All & Baidu Raven \\ \hline
Magenta & Hallo Magenta & Unclear & Unclear & Unclear & Part & Magenta \\ \bottomrule \hline
\end{tabularx}
\end{table*}

\begin{table*}[!htbp]
\centering
\caption{The information collected by the manufacturers.}
\label{real_world_info}
\scalebox{0.8}{ % 缩放比例
\begin{tabular}{l|cccccccccccccccccccc}
\toprule
Manufacturer & \multicolumn{20}{c}{Information Type} \\
\midrule
 & \rotatebox{90}{User Name and Contact Info} & \rotatebox{90}{User Email} & \rotatebox{90}{User Phone Number} & \rotatebox{90}{User Payment Info} & \rotatebox{90}{User Birth and Demographic Data} & \rotatebox{90}{User Coupons or Gift Card No.} & \rotatebox{90}{User Credit or Financing Info} & \rotatebox{90}{User Device Usage Data} & \rotatebox{90}{User App Usage Data} & \rotatebox{90}{User IP Address} & \rotatebox{90}{User Cookies Info} & \rotatebox{90}{User Service Usage Records} & \rotatebox{90}{User Device Type and Settings} & \rotatebox{90}{User Location Data} & \rotatebox{90}{User Voice Recordings and Transcripts} & \rotatebox{90}{User Social Media Account Info} & \rotatebox{90}{User Third-Party Site Info} & \rotatebox{90}{User Biometric Info} & \rotatebox{90}{User Comments or Posts} & \rotatebox{90}{Other User-Provided Info} \\
\hline
Amazon Alexa & \CheckmarkBold & \CheckmarkBold & \CheckmarkBold & \CheckmarkBold & \CheckmarkBold & \CheckmarkBold & \CheckmarkBold & \CheckmarkBold & \CheckmarkBold & \CheckmarkBold & \CheckmarkBold & \CheckmarkBold & \CheckmarkBold & \CheckmarkBold & \CheckmarkBold & \CheckmarkBold & \CheckmarkBold & \CheckmarkBold & \CheckmarkBold & \CheckmarkBold \\
Bixby & \CheckmarkBold & \CheckmarkBold & \CheckmarkBold & \CheckmarkBold & \CheckmarkBold & \CheckmarkBold & \CheckmarkBold & \CheckmarkBold & \CheckmarkBold & \CheckmarkBold & \CheckmarkBold & \CheckmarkBold & \CheckmarkBold & \CheckmarkBold & \CheckmarkBold & \CheckmarkBold & \CheckmarkBold & \CheckmarkBold & & \CheckmarkBold \\
Cortana & \CheckmarkBold & \CheckmarkBold & \CheckmarkBold & \CheckmarkBold & \CheckmarkBold & & & \CheckmarkBold & \CheckmarkBold & \CheckmarkBold & \CheckmarkBold & \CheckmarkBold & \CheckmarkBold & \CheckmarkBold & \CheckmarkBold & \CheckmarkBold & \CheckmarkBold & \CheckmarkBold & & \\
Clova & \CheckmarkBold & \CheckmarkBold & \CheckmarkBold & \CheckmarkBold & \CheckmarkBold & & & \CheckmarkBold & \CheckmarkBold & \CheckmarkBold & \CheckmarkBold & \CheckmarkBold & \CheckmarkBold & \CheckmarkBold & \CheckmarkBold & \CheckmarkBold & & \CheckmarkBold & & \\
Google Assistant & \CheckmarkBold & \CheckmarkBold & \CheckmarkBold & \CheckmarkBold & \CheckmarkBold & & & \CheckmarkBold & \CheckmarkBold & \CheckmarkBold & \CheckmarkBold & \CheckmarkBold & \CheckmarkBold & \CheckmarkBold & \CheckmarkBold & \CheckmarkBold & \CheckmarkBold & \CheckmarkBold & &  \\
Siri & \CheckmarkBold & \CheckmarkBold & \CheckmarkBold & \CheckmarkBold & \CheckmarkBold & & & \CheckmarkBold & \CheckmarkBold & \CheckmarkBold & \CheckmarkBold & \CheckmarkBold & \CheckmarkBold & \CheckmarkBold & \CheckmarkBold & \CheckmarkBold & & \CheckmarkBold & &  \\
Mycroft & \CheckmarkBold & \CheckmarkBold & \CheckmarkBold & & & & & \CheckmarkBold & \CheckmarkBold & \CheckmarkBold & \CheckmarkBold & \CheckmarkBold & \CheckmarkBold & \CheckmarkBold & \CheckmarkBold & & & \CheckmarkBold & \CheckmarkBold &  \\
Baidu Raven & \CheckmarkBold & \CheckmarkBold & \CheckmarkBold & \CheckmarkBold & \CheckmarkBold & & & \CheckmarkBold & \CheckmarkBold & \CheckmarkBold & \CheckmarkBold & \CheckmarkBold & \CheckmarkBold & \CheckmarkBold & \CheckmarkBold & \CheckmarkBold & \CheckmarkBold & \CheckmarkBold & & \\
Ding Dong & \CheckmarkBold & \CheckmarkBold & \CheckmarkBold & \CheckmarkBold & \CheckmarkBold & & & \CheckmarkBold & \CheckmarkBold & \CheckmarkBold & \CheckmarkBold & \CheckmarkBold & \CheckmarkBold & \CheckmarkBold & \CheckmarkBold & & & \CheckmarkBold & & \CheckmarkBold \\
Hallo Magenta & \CheckmarkBold & \CheckmarkBold & \CheckmarkBold & \CheckmarkBold & \CheckmarkBold & & & \CheckmarkBold & \CheckmarkBold & \CheckmarkBold & \CheckmarkBold & \CheckmarkBold & \CheckmarkBold & \CheckmarkBold & \CheckmarkBold & & & \CheckmarkBold & & \\
\bottomrule
\end{tabular}
}
\end{table*}

\section{Justification, Definitions and Descriptions in the Focus Group Stage}\label{sec:interview_definition}

We ran a non-judgmental focus group~\cite{moju2022you} where participants first generated ideas individually, then shared them aloud and finally discussed them collectively. The focus group's topic was ``what are the potentially influential factors affecting the risk perception of these data collections?''

We introduced the concepts of SPAs to participants as ``The intelligent personal assistant you would use in your home to complete daily tasks, control other devices, etc., as a hub.'' Privacy risks, often embedded in the environmental and social context, were introduced as ``potential negative outcomes such as data leakage, activity inference attacks or others you could think out.'' Data was introduced as ``information collected by SPAs during specific functions, with contextual significance, such as home location, rather than raw data like GPS coordinates.''

\section{Questionnaire Justifications and refinement}\label{sec:justification}

\textbf{Justification} The perceived risks of transmission and sharing may be influenced by the users' technological expertise and mental models regarding anonymity and encryption \cite{tabassum2019don}. As previous research has demonstrated, many users lack a comprehensive understanding of these data protection practices~\cite{li2023s,tabassum2019don}. Notably, these two factors also emerged in the interview process (see Section~\ref{sec:brainstorm_result}). To examine whether these factors mediate the observed effects, we included \textit{encryption} and \textit{anonymization} into our analysis. Specifically, encryption and anonymization are effective primarily when applied over the broad internet or within an organization, rather than at users' homes. Therefore, we associated encryption with \textit{transmitted across the broader internet} and anonymization with \textit{shared within company} and \textit{shared across third-party companies}. This approach aligns with the principles of CI by considering context-specific practices~\cite{nissenbaum2004privacy} (Figure~\ref{fig:transmis_share}).

We asked participants to assess privacy risks based on all potential disclosures and harms, as not all users were aware of every possible harm~\cite{abdi2019more,lau2018alexa}. We specifically focused on data sharing with companies (e.g., service providers or third-party companies) for three key reasons: 1. people were more willing to share data to non-commercial institutions (e.g., government \cite{roeber2015personal}, research institution \cite{roeber2015personal}) due to altruism \cite{skatova2014data,gaulin2020doing}. Such data-sharing scenarios are typically perceived as lower-risk~\cite{roeber2015personal}. 2. sharing or transmission of data to companies may be optional \cite{tikkinen2018eu}, as opposed to governments and regulatory institutions where disclosure may be mandatory, allowing users greater deciding agency. 3. users' mental models showed higher concerns of data sharing with companies (including third-party companies), which impacted their adoption~\cite{tabassum2019don}.

The measurement of \textit{awareness towards the data collection in correlated function} follows the guidance of the prior work~\cite{wickramasinghe2020survey}. \textbf{The importance of the data} was described as ``the significance of the data to complete the corresponding function. For example, location data may not be relevant for sending a WeChat message, but account data is essential.'' \textbf{Awareness of the data collection} was described as ``whether the participant is aware of the fact that the function collects the corresponding data''.

Due to the extensive data types (72 in total), we randomly selected 35-40 data types for each participant to rate. Each data type was rated 208 times in total. A Python algorithm using the shuffle function ensured even distributions of responses across data types. To reduce fatigue, we incorporated five evenly spaced 90-second breaks \cite{backor2007estimating} by inserting blank \textit{timer blocks}\footnote{Participants needed to wait until the time runs out to see the next questionnaire item.} into the questionnaire. This method ensured no participant dropped off the experiment.

\textbf{Pilot Studies} We conducted three rounds of pilot studies to refine the questionnaire. In the first round, we recruited 10 research assistants from XX university \footnote{anonymized for submission} to assess the clarity and comprehensibility of the items. This round also helped determine the appropriate number of entries for each participant to complete. In the second and third rounds, we recruited 10 participants from the same University to complete the questionnaire and identify any potential sources of misunderstanding. Issues raised in the second round were addressed, and no further concerns were reported during the third round. Therefore, the questionnaire was considered finalized and ready for deployment.

\section{Recruitment Details}\label{sec:study2_recruitment}

The quality controls of the questionnaire included 1) setting a minimum duration for questionnaire completion to be the number of questions times 5 seconds; 2) evenly embedding three attention-check questions in the questionnaire; 3) removing the outliers who consistently provided uniform responses (e.g., rate all entries with 7). A priori power analysis was conducted to determine the minimum sample size required for multiple linear regression analysis with 5 predictors (medium effect size: $f^2=0.07$, $\alpha=0.01$, $1-\beta=0.95$). The analysis indicated a required sample size of 402 participants. To account for potential attrition and ensure robust detection of smaller effects, we recruited 520 participants. The quality control methods filtered out 108 participants in total, reducing the sample from 520 to 412 participants, similar to previous studies~\cite{farke2021privacy}. 

\section{Interview Script}\label{sec:interview_script}

The following questions serve as the basis of the semi-structured interviews. Worth noting, to mitigate the gap between users' statements and their behavior, before and during the interview, we encouraged users to share their usage experience when answering the question.

\textbf{General Risk Perception:} 1. In your view, what are the key boundaries in your interaction with SPAs that help you assess privacy risks? For example, how do you distinguish between data that stays within the home versus data that moves outside your home network? 2. How do you categorize different types of data collected by SPAs (e.g., personal identifiers, location, health data)? What determines whether a particular type of data is more or less sensitive to you?

\textbf{Transmission and Sharing Concerns:} 3. What is your perception of the privacy risks when data is transmitted from your home network to the broad internet? Does the perceived risk change depending on whether it stays within the home network or goes beyond? 4. How do you assess the privacy risks of sharing your data with third-party companies? Does the entity collecting your data (e.g., the primary provider or a third party) influence your risk perception? Why?

\textbf{Encryption and Anonymization:} 5. Do you think encryption effectively mitigates privacy risks when data is transmitted beyond your home? How do you perceive its effectiveness compared to other privacy practices, such as anonymization? 6. Do you trust anonymization to protect your privacy when data is shared with third parties? How does anonymization affect your perception of risk in comparison to direct data transmission?

\textbf{Function-Specific Privacy Risks:} 7. How do you perceive the privacy risks associated with different SPA functions, such as health tracking, multimedia entertainment, or shopping? Are certain functions inherently riskier due to the data they require? 8. Does the type of data collected (e.g., location, health data, device usage) impact how you perceive privacy risks differently across various functions? How do you evaluate the sensitivity of data in the context of these functions?

\textbf{Boundary Regulation:} 9. How do you mentally map the flow of your data when using an SPA? Do you differentiate between the types of data that remain within your device, are shared with service providers, or are transmitted to third parties? 10. When you use an SPA, do you feel that your privacy boundaries are respected, or do you feel that some boundaries are being violated? What specific scenarios or actions make you feel that your privacy has been breached?

\section{Overall Ratings For Different Information Types Across Transmission and Sharing Ranges}

\begin{figure}[!htbp]
    \includegraphics[width=0.75\textwidth]{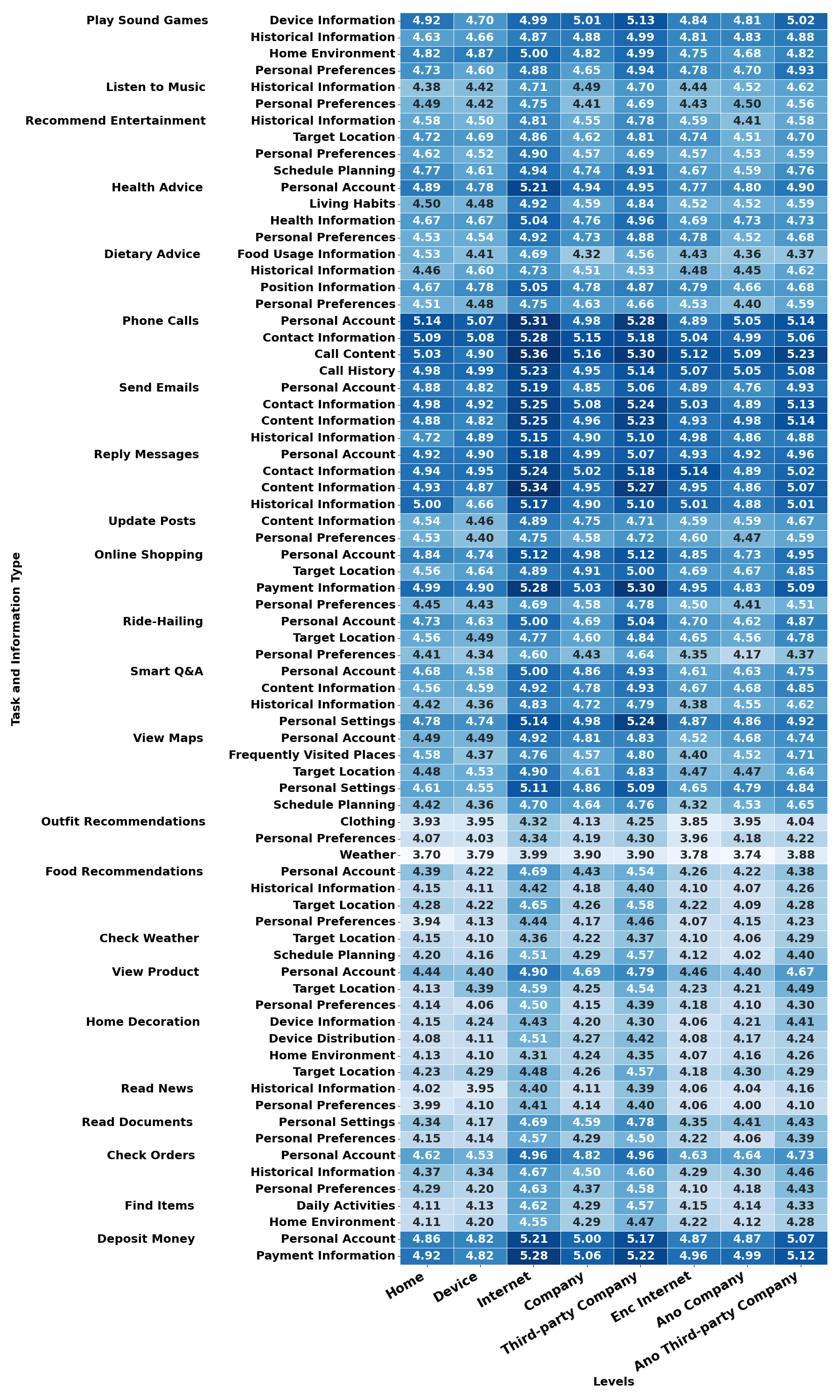}
    \caption{The perceived privacy risks of different information types across different transmission and sharing ranges.}
    \label{fig:information_type_rating}
\end{figure}

Figure~\ref{fig:information_type_rating} showed the overall ratings for different information types across transmission and sharing ranges, as well as encryption and anonymization settings. 

\end{document}